\documentclass[a4paper,12pt]{article}
\pdfoutput=1
\usepackage{amsmath}
\usepackage{amsfonts}
\usepackage{amssymb}
\usepackage{amstext}
\usepackage{graphicx}
\usepackage{mleftright}
\usepackage{epsfig}%
\title{\centerline \bf Evolution of scalar and vector cosmological
  perturbations through a bounce in metric $f(R)$ gravity in flat FLRW
  spacetime}
\bigskip
\author{Pritha Bari$^\dagger$, Kaushik Bhattacharya$^\ddagger$
  \thanks{$^\dagger$ pribari@iitk.ac.in,
  $^\ddagger$kaushikb@iitk.ac.in}\\
\normalsize Department of Physics, Indian Institute of
Technology, Kanpur\\
\normalsize Kanpur 208016, India }
\begin{document}
\maketitle
\begin{abstract}
In the present work we present the full treatment of scalar and vector
cosmological perturbations in a non-singular bouncing universe in the
context of metric $f(R)$ cosmology. Scalar metric perturbations in
$f(R)$ cosmology were previously calculated in the Jordan frame, in
the present paper we successfully use the Einstein frame to calculate
the scalar metric perturbations where the cosmological bounce takes
place in the Jordan frame. The Einstein frame picture presented
corrects and completes a previous calculation of scalar perturbations and
adds new information. Behavior of fluid velocity potential and the
pure vector fluid velocity terms are elaborately calculated for the
first time in $f(R)$ cosmology for a bouncing universe in presence of
exponential gravity. It is shown that the vector perturbations can
remain bounded and almost constant during the non-singular bounce in
$f(R)$ gravity unlike general relativistic models where we expect the
vector perturbations to be growing during the contracting phase and
decaying during the expanding phase. The paper shows that the Einstein
frame can be used for calculation of scalar and vector metric
perturbations in a bouncing universe for most of the cases except the
case of asymmetric non-singular bounces.
\end{abstract}
\section{Introduction}

Though inflation\cite{Starobinsky:1980te,Guth:1982ec} has been
tremendously successful in solving most of the problems of the
standard big bang cosmology, the issue of singularity and
trans-Planckian problem still remains
\cite{Borde:1996pt,Martin:2000xs,Brandenberger:2012aj}. A possible
solution to the above mentioned problems is to consider the existence
of non-singular bouncing cosmologies
\cite{Martin:2001ue,Martin:2003sf,Battefeld:2014uga,
  Novello:2008ra,Cai:2012va}, in which the universe goes from a
contracting phase to an expanding one through bounce without any
singularity. Non-singular bounce also addresses the horizon and
flatness problem\cite{Battefeld:2014uga}. If one does not want to
introduce some exotic matter components in a 3-dimensionally flat
Friedmann-Lemaitre-Robertson-Walker (FLRW) spacetime then a way of
realizing bounce is modifying general relativity (GR)
\cite{Abramo:2009qk,Carloni:2005ii,Paul:2014cxa,Bhattacharya:2015nda,Bamba:2013fha,
  Bari:2018aac}. It is conjectured that GR may not be the unique,
correct theory of gravity to describe geometry of space-time when the
curvature scale is high. Modifications of the Einstein-Hilbert
gravitational action by higher order curvature invariants is done in
very strong gravity regimes, such as in very early universe. At high
curvature limits, when bounce happens, modifications to GR is
expected. In this paper we study perturbations in a bouncing cosmology
where the theory of gravity is given by metric $f(R)$ theory. The
$f(R)$ paradigm is important as various models of inflation and late
time acceleration of the universe can be modelled on $f(R)$ gravity
theories. Current observation has detected a cosmic acceleration
starting after the matter domination. Modified gravity theories
\cite{Myrzakulov:2013hca,Clifton:2011jh,Atazadeh:2006re,Carroll:2003wy}
have been used to explain this late time acceleration of the universe.
Although $f(R)$ gravity \cite{Nojiri:2017ncd} \cite{Nojiri:2010wj} is
only one amongst the many modified gravity models, but it is one of
the simplest modifications of GR  which can tackle various cosmological problems.

It is known that $f(R)$ theory of gravity can be analyzed
\cite{Paul:2014cxa,Maeda:1988ab,Sotiriou:2008rp,DeFelice:2010aj} in two conformal
frames; the Jordan frame and the Einstein frame.  In Jordan frame the
theory is a higher derivative theory, because higher than two order of
time derivatives appear in the field equation. The Jordan frame field
equation is obtained by varying the $f(R)$ action with respect to the
metric tensor \cite{Sotiriou:2008rp}. In $f(R)$ cosmology is is
assumed that the problem of cosmological dynamics is fundamentally
posed in the Jordan frame although one can work the cosmological
dynamics also in the Einstein frame using the conformal transformation
connecting the frames. An advantage of working in the Einstein frame
is that the theory of gravity becomes GR and known techniques of GR
evolution can be applied in the Einstein frame. One can always
transform back to the Jordan frame after calculating cosmological
dynamics in the Einstein frame. In previous studies this was the
method followed \cite{Paul:2014cxa}. Einstein frame description of
$f(R)$ is GR with an added scalar field that is minimally coupled to
gravity and non-minimally coupled to matter. Einstein frame description
of $f(R)$ gravity provides easier ways to tackle many problems in
$f(R)$ gravity.

There have been many attempts to realize bounce in various $f(R)$
gravity models \cite{Bamba:2013fha} \cite{Carloni:2005ii}. It was
pointed out in \cite{Bari:2018aac} that exponential $f(R)$ theory
might be a good candidate theory which supports a bounce in flat
FLRW metric. We will, hence, deal with exponential $f(R)$ gravity in
this paper. The present work deals with scalar and vector
perturbations in a bouncing universe guided by exponential gravity,
the bounce takes place in the Jordan frame. The perturbations have
been calculated in both the Jordan and Einstein frames and then the
results are matched to gain insight into the nature of the conformal
correspondence of the two frames. It has been shown that the Einstein
frame can be used for the calculation of scalar cosmological perturbations
for symmetric bounce in the Jordan frame. The conformal correspondence
fails for asymmetrical bounces in the Jordan frame. No such
difficulties arise for vector perturbations where the Einstein frame
can be safely used for all the calculations in a much easier way. A
part of scalar metric perturbation calculation in the Einstein frame
was incompletely presented in an earlier work Ref.~\cite{Paul:2014cxa}
where the authors did not take into account the role of fluid velocity
potential. In the present work we specify the complete and correct way
of calculation of the scalar metric perturbations in a bouncing
universe using the Einstein frame. We present the full nature of the
scalar perturbations and the dynamics of the fluid velocity potential
during a cosmological bounce. The next part of the paper shows the
nature of vector metric perturbations during a non-singular bounce. In
$f(R)$ cosmology the vector perturbations remains almost constant
during the bounce which differs from GR results where the vector
perturbations generally increases during the contraction phase
\cite{Battefeld:2004cd}. Throughout the paper the role of the Einstein frame as an
important frame for calculations has been emphasized. The issue about tensor
perturbations in a bouncing universe will be addressed in a future work.

The material in the paper is presented in the following manner. The
next section presents the background cosmological evolution in both
the Jordan frame and Einstein frame. It also specifies the conformal
transformation relating these frames. Section \ref{scalar} introduces
the scalar metric perturbations in both the frames. The Jordan frame
result is first calculated and next the Einstein frame results are
presented. The results of scalar perturbations are calculated for both
a symmetric and asymmetric bounce in the Jordan frame.  The topic of
vector perturbations is taken up in section \ref{vp}. The next and the
last section concludes the present work with a brief summary of the
results obtained.
\section{Field equations of $f(R)$ gravity} 
\label{back}

In the following we work with the spatially flat maximally symmetric
FLRW metric given by
\begin{equation}
ds^2=-dt^2+a^2(t)d{\bf x}^2
\label{g}
\end{equation}
where $t$ is the cosmological time, ${\bf x}$ is the Co-moving spatial
coordinates and $a(t)$ is the scale-factor of the universe. In the
domain of general relativity it is well known that a bouncing solution
is possible only for spatially positively curved FLRW universe, if we
do not want to include any exotic matter component in the
scenario. But in metric $f(R)$ gravity and as shown in some previous
works\cite{Paul:2014cxa,Bhattacharya:2015nda}, it is possible to have
bouncing solution in spatially flat FLRW universe without invoking any
exotic matter component for certain $f(R)$ theories, simplest of which
being the $R+\alpha R^2$ gravity with a negative $\alpha$. The reason
is that the extra curvature induced energy density and pressure terms
can indeed produce the bouncing conditions. In this section we present
the field equations in Jordan and Einstein frame respectively.  We
will later apply the formalism to study classical cosmological bounce
phenomena \cite{Novello:2008ra}.
\subsection{The relevant cosmological equations in Jordan and Einstein
  frames}
\label{jefr}
The modified Friedmann equations for $f(R)$ theory in the Jordan frame are given as
\cite{Sotiriou:2008rp}:
\begin{eqnarray}
3 H^2 &=&\frac{\kappa}{F(R)} \rho_{\rm eff}\,,
\label{fried}\\
3H^{2}+2\dot{H} &=&-\frac{\kappa}{F(R)} P_{\rm eff}\,,
\label{2ndeqn}
\end{eqnarray}
where $H$ is the conventional Hubble parameter defined as $H\equiv
\dot{a}/a$ and the constant $\kappa = 8\pi G$
where $G$ is the universal gravitational constant. In the above equations
\begin{eqnarray}
F(R) \equiv \frac{df(R)}{dR}\,.
\label{fp}
\end{eqnarray}
The dot specifies a derivative with respect to cosmological time $t$.
The effective energy density, $\rho_{\rm eff}$, and pressure, $P_{\rm
  eff}$, are defined as:
\begin{eqnarray}
\rho_{\rm eff} \equiv \rho + \rho _{\rm curv}\,,\,\,\,\,\,
P_{\rm eff} \equiv P + P_{\rm curv}\,,
\label{epeff}
\end{eqnarray}
where $\rho_{\rm curv}$ and $P_{\rm curv}$ are given by
\begin{eqnarray}
\rho _{\rm curv} &\equiv& \frac{RF-f}{2\kappa}-\frac{3H\dot{R}
F_R(R)}{\kappa}\,,
\label{reff}\\
P_{\rm curv} &\equiv& \frac{\dot{R}^{2}F_{RR} + 2H\dot{R}F_R
  + \ddot{R}F_R }{\kappa} - \frac{RF-f}{2\kappa}\,,
\label{peff}
\end{eqnarray}
which are curvature induced energy-density and pressure. In the above
equations the subscript $R$ specifies derivatives with respect to the
Ricci scalar $R$. The curvature induced thermodynamic variables exists
in absence of any hydrodynamic matter. The conventional
$\rho$ and $P$ are defined through
\begin{eqnarray}\label{tmunu}
T_{\mu \nu} = (\rho + P)u_\mu u_\nu + P g_{\mu\nu}\,, 
\end{eqnarray}
which has the information of hydrodynamic matter. In this article we
assume the fluid to be barotropic so that its equation of state is
\begin{eqnarray}
P=\omega \rho\,,
\label{eqns}
\end{eqnarray}
where $\omega$ is a constant and its value is zero for dust and
one-third for radiation.  It must be noted that $u_\mu$ in
Eq.~(\ref{tmunu}) is the 4-velocity of a fluid element and $u_\mu
u^\mu = -1\,.$ 

One can make a conformal transformation on the system of equations in
the Jordan frame to recast the problem in the Einstein frame. The
Einstein frame version of the cosmological dynamics sometimes becomes
relatively easy to manage as in this version one deals with the known
Einstein equations. The Einstein frame description of $f(R)$ gravity
is obtained by the following conformal transformation,
\begin{eqnarray}
\tilde{g}_{\mu \nu}=F(R)g_{\mu \nu}\,,
\label{gtilde}
\end{eqnarray}
and simultaneously defining a new scalar field $\phi$ as
\begin{eqnarray}
\phi = \sqrt{\frac{3}{2\kappa}} \ln F(R)\,.
\label{phidef}
\end{eqnarray}
This scalar field plays an important role in the Einstein frame.  The
conformally transformed line element in the Einstein frame is
\begin{eqnarray} 
d\tilde{s}^2 = -d\tilde{t}^2 + \tilde{a}^2 d{\bf x}^2\,,
\label{dseins}
\end{eqnarray}
where the time coordinate, $\tilde{t}$, and the scale factor,
$\tilde{a}$, in the Einstein frame are related to their corresponding
Jordan frame terms via the relations
\begin{eqnarray}
d\tilde{t}= \sqrt{F(R)} \,dt\,\,\,\,\,{\rm and}\,\,\,\,\,
\tilde{a}(t)=\sqrt{F(R)} \,a(t)\,.
\label{aat}
\end{eqnarray}
Using these transformations one can formulate the gravitational
dynamics of $f(R)$ gravity in the Einstein frame in presence of matter
and the scalar field $\phi$ acting as sources. The energy-momentum
tensor in the Einstein frame, which is related to $T^{\mu \nu}$ in the
Jordan frame, turns out to be
\begin{eqnarray}
\tilde{T}_{\mu \nu} = (\tilde{\rho} +
\tilde{P})\tilde{u}_\mu \tilde{u}_\nu + \tilde{P} \tilde{g}_{\mu\nu}\,,
\label{tmnt}
\end{eqnarray}
where $\tilde{\rho}=\rho/F^2(R)$, $\tilde{P}=P/F^2(R)$ and $\tilde{u}_\mu =
\sqrt{F(R)}u_\mu$. In the Einstein frame
$\tilde{g}^{\mu\nu}\tilde{u}_\mu \tilde{u}_\nu=-1$.
Except $\tilde{T}^{\mu \nu}$, the energy-momentum tensor for the scalar field 
also acts as source of curvature in the Einstein frame and it is given as
\begin{eqnarray}
S^{\mu}_{\nu}=\partial_\alpha \phi \partial_\nu \phi\tilde{g}^{\alpha
  \mu} - \delta^\mu_\nu {\mathcal L}(\phi)\,, 
\label{smn}
\end{eqnarray}
where the scalar field Lagrangian is
\begin{eqnarray}
{\mathcal L}(\phi)= \frac12 \partial_\alpha \phi \partial_\beta\phi 
\tilde{g}^{\alpha \beta} + V(\phi)\,.
\label{philag}
\end{eqnarray}
The scalar field potential in the Einstein frame turns out to be
\begin{eqnarray}
V(\phi)=\frac{RF-f}{2\kappa F^2}\,,
\label{potphi}
\end{eqnarray}
where one has to express $R=R(\phi)$, from Eq.~(\ref{phidef}) by
inverting it, and then express $V(\phi)$ as an explicit function of
$\phi$. From the form of $S^{\mu}_{\nu}$ one can write
\begin{eqnarray}
S^0_{0} \equiv \rho_\phi = \frac12 \left(\frac{d\phi}{d\tilde{t}}\right)^2
+ V(\phi)\,,\,\,\,\,
S^i_{i} \equiv P_\phi = \frac12 \left(\frac{d\phi}{d\tilde{t}}\right)^2
- V(\phi)\,,
\label{rps}
\end{eqnarray}
where the scalar field $\phi$ is assumed to be a function of time
only. The total energy-momentum tensor responsible for gravitational
effects in the Einstein frame is $\tilde{T}^\mu_{\,\,\,\,\nu} + S^{\mu}_{\,\,\,\,\nu}$
which is a mixed tensor with only diagonal components. 

The time coordinate, $t$, and the Hubble parameter, $H$, in the the
Jordan frame are related to the time coordinate, $\tilde{t}$, and
Hubble parameter $\tilde{H}(\equiv
\frac{1}{\tilde{a}}\frac{d\tilde{a}}{d\tilde{t}})$, in 
the Einstein frame via the relations:
\begin{eqnarray}
\tilde{t}=\int_{t_0}^t \sqrt{F(R)} dt'\,,\,\,\,\,\,\,\,
H=\sqrt{F}\left(\tilde{H}-\sqrt{\frac{\kappa}{6}}\,\frac{d\phi}{d\tilde{t}}
\right)\,.
\label{th}
\end{eqnarray}
As we will be interested mainly in bouncing cosmologies $t_0$ will be
set to zero. The instant $t_0=0$ is the bouncing time in the Jordan
frame.  The Einstein frame description of cosmology can be tackled
like FLRW spacetime in presence of a fluid and a scalar field. The
presence of the Scalar field potential $V(\phi)$ gives one a pictorial
understanding of the physical system which is lacking in the Jordan
frame. Seeing the nature of the potential and the initial conditions
of the problem one gets a hint about the possible time development of
the system.  The time evolution of the scalar field in the Einstein
frame is dictated by the equation
\begin{eqnarray}
\frac{d^2 \phi}{d\tilde{t}^2} + 3\tilde{H}\frac{d\phi}{d\tilde{t}}
+\frac{dV}{d\phi}=\sqrt{\frac{\kappa}{6}}(1-3\omega)\tilde{\rho}\,,
\label{phieqn}
\end{eqnarray}
where the equation of state of the fluid in the Jordan frame is
$P=\omega \rho$. It is interesting to note that the equation of state
of the fluid remains the same in the Einstein frame. The evolution of
the energy density in the Einstein frame is given by
\begin{eqnarray}
\frac{d\tilde{\rho}}{d\tilde{t}}+\sqrt{\frac{\kappa}{6}}(1-3\omega)\tilde{\rho}
\frac{d \phi}{d\tilde{t}} + 3\tilde{H} \tilde{\rho}(1+\omega)=0\,.
\label{rhotildeq}
\end{eqnarray}
The above two equations dictate the time evolution of $\tilde{\rho}$
and $\phi$ in the Einstein frame. To generate proper bouncing solution
from the above two equations one requires the values of only two
quantities at the bouncing time, they are $\phi(\tilde{t}_0)$,
$\left(\frac{d\phi}{d\tilde{t}}\right)_{\tilde{t}_0}$, the other
parameters are determined from these two at the bouncing
time\cite{Paul:2014cxa}. In general for a flat FLRW spacetime these
two values of the respective quantities are enough to solve the whole
system in the Einstein frame. The expression of the Hubble parameter
and its rate of change in the Einstein frame are given as
\begin{eqnarray}
\tilde{H}^2 &=& \frac{\kappa}{3}(\rho_\phi + 
\tilde{\rho})\,,
\label{htilde}\\
\frac{d\tilde{H}}{d\tilde{t}}&=&-\frac{\kappa}{2}
\left[\left(\frac{d\phi}{d\tilde{t}}\right)^2 + (1+\omega) \tilde{\rho}\right]\,.
\label{hprime}
\end{eqnarray}

\section{Scalar cosmological perturbations in $f(R)$ gravity}
\label{scalar}

Scalar perturbation in the cosmological framework, mainly related to
inflation, has been widely studied \cite{Riotto:2002yw,
  Baumann:2009ds, Mukhanov:1990me,Xue:2013bva}.  In this section we
discuss the evolution of scalar cosmological perturbation through a
bounce in $f(R)$ gravity both in Jordan frame and Einstein frame.  For
the sake of comparison, it is better to use the conformal time $\eta$,
since by definition it is invariant under a conformal transformation,
$d\eta={dt}/{a(t)}={d\tilde{t}}/{\tilde{a}(\tilde{t})}$.  The most
general form of the line element in the Jordan frame is:
\begin{equation}
  ds^2= a(\eta)^2[-(1+2 \chi) d\eta^2+2B_{,i} dx^i d\eta + ((1-2\psi)\delta_{ij}+
  2E_{,ij})dx^i dx^j]\,.
\end{equation}
Here $\chi$, $B$, $\psi$ and $E$ are functions of space and time. The
subscript(s) preceded by a comma specifies partial derivatives.
The corresponding expression in Einstein frame for the perturbed spacetimes is:
\begin{equation}
  d\tilde{s}^2= (F+\delta F) ds^2 = \tilde{a}(\eta)^2[-(1+2 \tilde{\chi}) d\eta^2 + 2\tilde{B}_{,i} dx^i d\eta + ((1-2\tilde{\psi})\delta_{ij}+2\tilde{E}_{,ij})dx^i dx^j],
\end{equation}
where henceforth $F$ stands for the unperturbed value of $df/dR$. The
scale-factors are still related via the relation specified in
Eq.~(\ref{aat}). The scalar metric perturbation functions in the two
frames are related via \cite{DeFelice:2010aj},
\begin{align}
  \tilde{\chi}= \chi + \frac{\delta F}{2F}\,, && {\rm and} && \tilde{\psi}=
  \psi - \frac{\delta F}{2F}\,,
\label{chipsi}
\end{align}
\begin{align}
\tilde{B}= B\,, && {\rm and} && \tilde{E}= E\,.
\label{be}
\end{align}
In Jordan frame the gauge-invariant
variables are:
\begin{equation}
\Phi= \chi + \frac{1}{a}[(B-E')a)]'\,,\,\,\,\,
\Psi= \psi - \frac{a'}{a}(B-E')\,. 
\end{equation}
The corresponding gauge-invariant variables, in the Einstein frame, are:
\begin{equation}
  \tilde{\Phi}= \tilde{\chi} + \frac{1}{\tilde{a}}[(\tilde{B}-\tilde{E}')\tilde{a})]'
  = \Phi + \frac{\delta F^{(gi)}}{2F}\,,\,\,\,\,\,
  \tilde{\Psi}= \tilde{\psi} - \frac{\tilde{a}'}{\tilde{a}}(\tilde{B}-\tilde{E}') =
  \Psi - \frac{\delta F^{(gi)}}{2F}\,,
\end{equation}
where $\delta F^{(gi)}=(\partial F/\partial R)\delta R^{(gi)}$. The
gauge invariant perturbation of a scalar quantity $q$ is
\cite{Mukhanov:1990me}, $\delta q^{(gi)}\equiv\delta
q+q^\prime(B-E^\prime)\,.$ Here the primes stand for derivatives with
respect to conformal time.  Assuming the perturbation in the matter
sector in Jordan frame to be such that
\begin{equation}
\delta P^{(gi)}-c_{s}^{2}\delta\rho^{(gi)}=0\,,
\end{equation}
where $c_s^2={\delta P}/{\delta\rho}$ one can derive the relation
between $\tilde{P}^{(gi)}$ and $\tilde{\rho}^{(gi)}$ in the Einstein
frame as
\begin{eqnarray}
  \delta\tilde{P}^{(gi)}-c_{s}^{2}\delta\tilde{\rho}^{(gi)}&=&\delta \tilde{P}-
  c_s^2\delta\tilde{\rho}\nonumber\\
  &=&\delta(Pe^{-2\sqrt{\frac{2\kappa}{3}}\phi})-c_s^2\delta(\rho e^{-2\sqrt{\frac{2\kappa}{3}}\phi})
  \nonumber\\
  &=&e^{-2\sqrt{\frac{2\kappa}{3}}\phi}(\delta P-c_s^2\delta\rho)+2\sqrt{\frac{2\kappa}{3}}
  e^{-2\sqrt{\frac{2\kappa}{3}}\phi}(c_s^2\rho-P)\delta\phi\nonumber\\
  &=&e^{-2\sqrt{\frac{2\kappa}{3}}\phi}(\delta P^{(gi)}-c_s^2\delta\rho^{(gi)})+
  2\sqrt{\frac{2\kappa}{3}}e^{-2\sqrt{\frac{2\kappa}{3}}\phi}(c_s^2\rho-P)\delta\phi
  \nonumber\\
&=&2\sqrt{\frac{2\kappa}{3}}\tilde{\rho}(c_{s}^{2}-\omega)\delta\phi\,,
\end{eqnarray}
showing that for a single barotropic fluid (where $\omega=c_s^2$)
whose equation of state is given by Eq.~(\ref{eqns}) in the Jordan
frame one must have $\delta\tilde{P}^{(gi)}-c_{s}^{2}\delta\tilde{\rho}^{(gi)}=0$ in the
Einstein frame.

Before we proceed to formulate the scalar cosmological perturbation in
the Jordan and Einstein frames it is pertinent to elucidate the nature
of gauge choices and their relationship with conformal
transformations.  If we apply synchronous gauge in Jordan frame, one
should impose $\chi=B=0$. It can be noted that although
$\tilde{B}=B=0$ in Einstein frame, $\tilde{\chi}$ does not
vanish. So the definition of synchronous gauge is not the same in both
frames.

In the spatially-flat gauge, in Jordan frame, one has to impose
$\chi=E=0$.  From Eq.~(\ref{chipsi}) it is seen that $\tilde{\chi}$
does not vanish in Einstein frame. Consequently the spatially-flat gauge is also not conformally invariant.

Only in the longitudinal gauge, where one imposes $B=E=0$ in the
Jordan frame, one obtains $\tilde{B}=\tilde{E}=0$ in Einstein
frame. The longitudinal gauge remains invariant under the conformal
transformation connecting the Jordan frame and the Einstein frame.
Hence we will use longitudinal gauge in the present article from here
on.
\subsection{Scalar perturbations in the Jordan frame}
\label{jords}

In the domain of linear perturbations, the scalar perturbed FLRW
metric has two gauge invariant degrees of freedom. In the longitudinal
gauge this can be expressed as,
\begin{eqnarray}
ds^{2}=a^{2}(\eta)\left[ -(1+2\Phi)d\eta^{2}+(1-2\Psi)\delta_{ij} 
dx^{i}dx^{j}\right]
\label{delta_g}
\end{eqnarray}
Here $\Phi$ and $\Psi$ are the two gauge invariant perturbation
degrees of freedom, also called the Bardeen potentials. The, $00$,
$ii$, $ij-th(i\neq j)$ elements of the linearized perturbed Einstein
equation in the Fourier space are \cite{Matsumoto:2013sba}:
\begin{eqnarray}
  F[-k^2(\Phi+\Psi)-3\mathcal{H}(\Phi^\prime+\Psi^\prime)+(3\mathcal{H}^\prime -6\mathcal{H}^2)
    \Phi-3\mathcal{H}^\prime\Psi]\nonumber\\
+F^\prime(-9\mathcal{H}\Phi+3\mathcal{H}\Psi-3\Psi^\prime)=\kappa a^2\delta\rho\,, 
\label{jord_ptbd_00}
\\
\nonumber\\
F[\Phi^{\prime\prime}+3\mathcal{H}(\Phi^\prime+\psi^\prime)+3\mathcal{H}^\prime
  +(\mathcal{H}^\prime+2\mathcal{H}^2)\Psi]+F^\prime(3\mathcal{H}\Phi-\mathcal{H}\Psi+
3\Phi^\prime)\nonumber\\
+F^{\prime\prime}(3\Phi-\Psi)=c_s^2\kappa a^2\delta P \,,
\label{jord_ptbd_ii}
\\
\nonumber\\
\Phi-\Psi-\frac{2F_R}{a^2 F}[3\Psi^{\prime\prime}+6(\mathcal{H}^\prime+\mathcal{H}^2)\Phi
  +3\mathcal{H}(\Phi^\prime+3\Psi^\prime)-k^2(\Phi-2\Psi)]=0 \,,
\label{jord_ptbd_ij}
\end{eqnarray}
where $F_R=dF/dR$. If there is only a single matter component present, the perturbation
in the matter sector can be assumed to be adiabatic, so that the sound
velocity can be defined as done in the beginning of this section.
Using Eq.~(\ref{jord_ptbd_00}) and (\ref{jord_ptbd_ii}), we can write
\begin{figure}[t!]
\centering
\includegraphics[scale=.5]{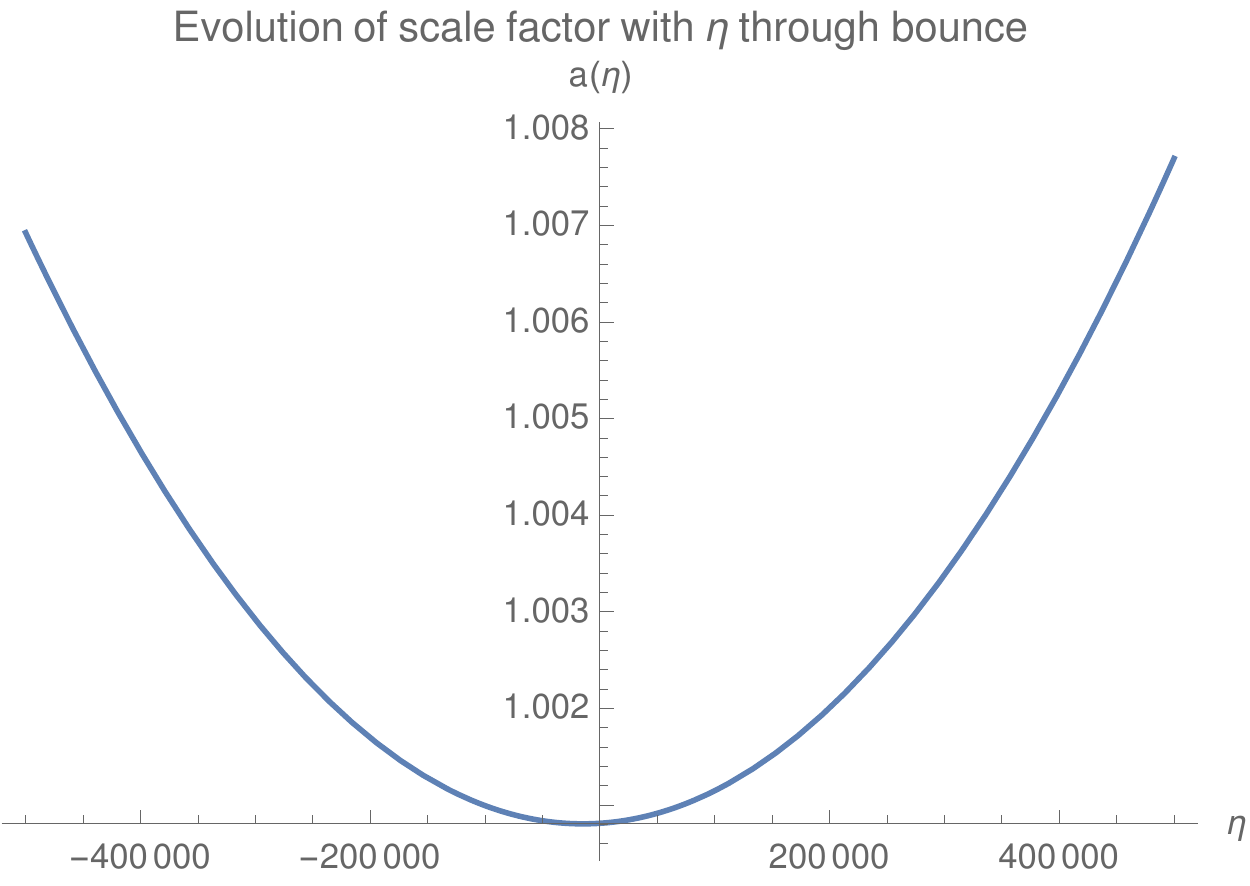}
\caption{Evolution of scale factor with $\eta$ through asymmetric bounce} 
\label{bp2}
\end{figure}
\begin{eqnarray}
&&(1+F)\left[\Phi^{\prime\prime}+\Psi^{\prime\prime}+3\mathcal{H}(1+c_s^2)
(\Phi^\prime + \Psi^\prime)+c_s^2(k^2+6\mathcal{H}^2)\Phi+3\mathcal{H}^\prime(1-c_s^2)\Phi+
\mathcal{H}^\prime(1+c_s^2)\Psi\right.\nonumber\\
&&+\left.(2\mathcal{H}^2+c_s^2k^2)\Psi\right]+F^\prime[\mathcal{H}(1+3c_s^2)(\Phi-3\Psi)+
3\Phi^\prime + 3c_s^2\Psi^\prime]+F^{\prime\prime}(3\Phi-\Psi)=0\,.
\label{jord_ptbd_ii00}
\end{eqnarray}
A form of the above perturbation equations of $f(R)$ gravity in the
Jordan frame was also calculated in \cite{Bean:2006up} where the
authors studied the problem of structure formation in late times. In
this paper we will follow the equations as written above and obtained
from \cite{Matsumoto:2013sba} which seem more appropriate for our
analysis. Eq.~(\ref{jord_ptbd_ij}) and (\ref{jord_ptbd_ii00}) can be
solved numerically to get the solutions $\Phi$ and $\Psi$.  In the
present article we will exclusively work with exponential gravity
\cite{Bari:2018aac} where the form of $f(R)$ is given as
\begin{equation}
f(R)= \frac{1}{\alpha} \exp{(\alpha R)}\,,
\label{frform}  
\end{equation}
where $\alpha= 10^{12}$ for phenomenological reasons, it sets the
energy scale of bounce\footnote{The word exponential gravity in the
  context of $f(R)$ theories may be a bit confusing as previously many
  authors have used exactly the same name to work a completely
  different problem. The authors of the works \cite{Bean:2006up,
    Linder:2009jz, Odintsov:2017qif, Bamba:2010ws} have used
  exponential gravity to tackle late time universe problems as
  structure formation or the dark energy problem. In all of these
  cases the form of $f(R)$ is given as $R + g(R)$ where the function
  $g(R)$ contains an exponential function of the Ricci
  scalar. Compared to those models our model is much simpler and more
  over our model of exponential gravity tackles an early universe
  problem. As the form of $f(R)$ we are using is purely exponential we
  do not modify the name of the theory.}. In the present work all the
values of dimensional constants (as $\alpha$) or variables (as ${\cal
  H}$, $\eta$, and others) will be represented in Planck units. To go
back to mass units one has to multiply the appropriate variables by
suitable power of Planck mass expressed in GeV units. We choose
exponential gravity as in this case the gravitational theory does not
have any instabilities, as because with $\alpha > 0$ we have
$$F>0\,,\,\,\,\,\,F_R>0\,.$$ The issues about instability in this kind of theory is 
discussed in \cite{Bari:2018aac}.
\begin{figure}[t!]
\begin{minipage}[b]{0.5\linewidth}
\centering
\includegraphics[scale=.5]{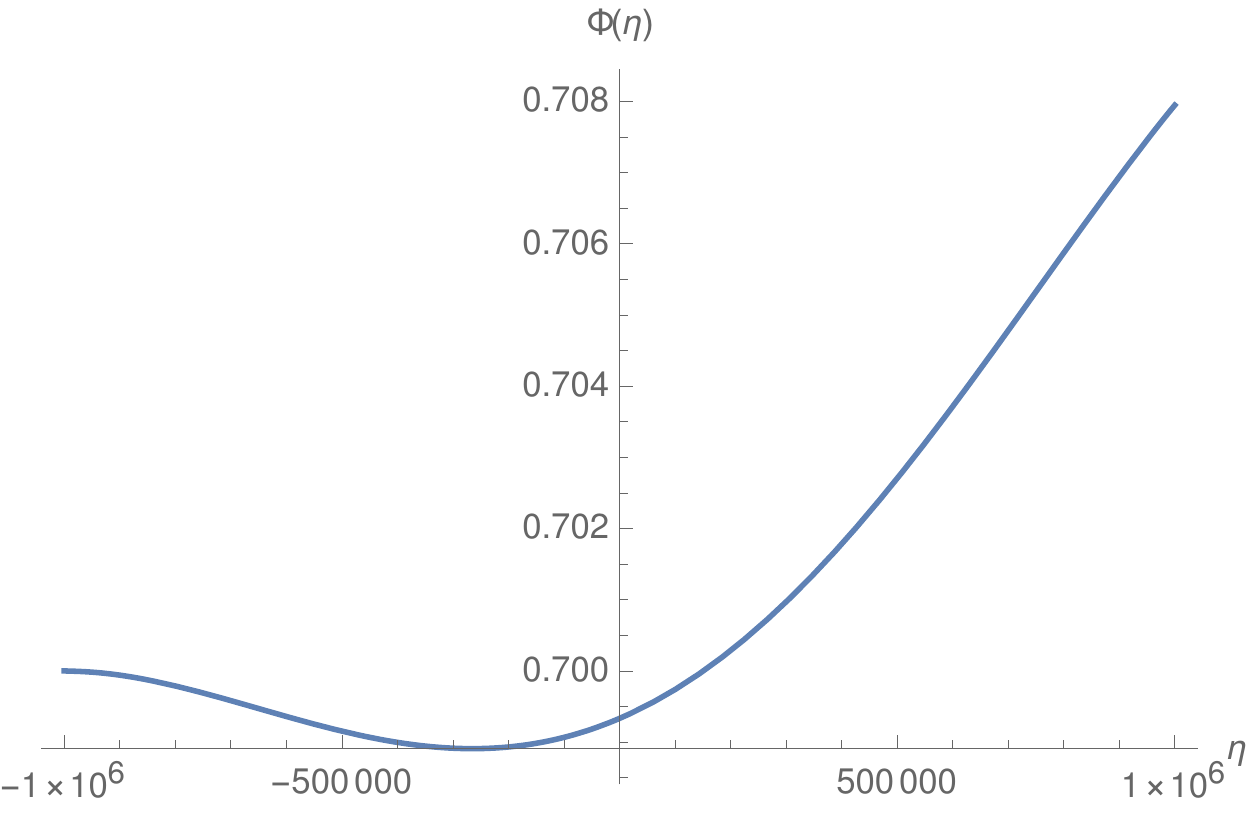}
\caption{Evolution of $\Phi$ with $\eta$ through asymmetric bounce, calculated directly in Jordan frame} 
\label{pb}
\end{minipage}
\hspace{0.2cm}
\begin{minipage}[b]{0.5\linewidth}
\centering
\includegraphics[scale=.5]{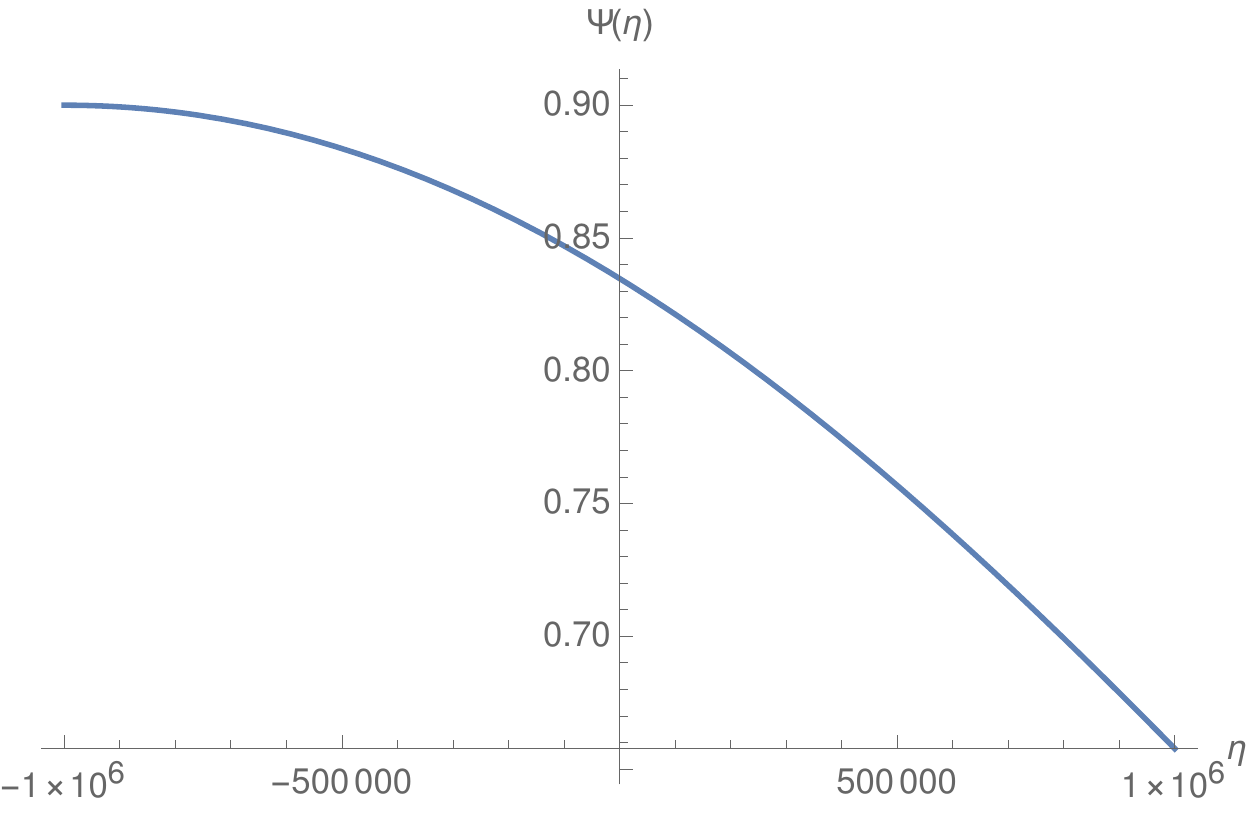}
\caption{Evolution of $\Psi$ with $\eta$ through asymmetric bounce, calculated directly in Jordan frame}
\label{pa}
\end{minipage}
\end{figure}

The blue curves in Fig.~\ref{pn2} and Fig.~\ref{pn3} shows the
evolution of $\Phi(k,\eta)$ and $\Psi(k,\eta)$ for $k=10^{-10}$ for a
particular background evolution. We have used $\Phi(0)=0.0001$,
$\Phi'(0)=0, \Psi(0)=0.0002, \Psi'(0)=0$ to produce the plots. The
bouncing time is $\eta=0$. For the background we have chosen radiation
where $\omega=1/3$. The bounce in the background is shown in
Fig.~\ref{bp2} where the scale-factor is plotted with respect to
$\eta$.  To obtain the the background solutions we have initially
solved the system of equations as specified in the initial part of
section \ref{back} in Jordan frame. The analysis is done in coordinate
time. Later the result is transformed to conformal time.  The
conditions used to produce the bouncing background solution are,
${\cal H}(0)=0$, ${\cal H}^\prime(0)=6.8 \times 10^{-14}$ and ${\cal
  H}^{\prime \prime} = 0$.  The background evolution in Fig.~\ref{bp2}
specifies a symmetric bounce. In all the calculations of bounce we
have normalized the scale-factor in the Jordan frame in such a way
that $a(0)=1$.  One can also choose the bounce in the background to be
asymmetric. In this case the conditions at $\eta=0$ remains the same
as before (as in the symmetric case) except that ${\cal H}^{\prime
  \prime}(0) \ne 0$. The asymmetry in the background evolution can be
generated from an infinitesimal value of the second time derivative of
the Hubble parameter as ${\cal H}^{\prime \prime}(0)=8.2 \times
10^{-21}$. The perturbation evolution in an asymmetric background in
the Jordan frame are plotted in Fig.~\ref{pb} and Fig.~\ref{pa}. To
evaluate the dynamics of the perturbations the values of the
perturbations are now not specified at $\eta=0$ but at $\eta=-10^6$.
The reason for choosing a separate time instant for plotting the
asymmetric bounce results will become clear when we analyze the case
of asymmetric bounces in the Einstein frame. In the present case
$\Phi(-10^6)=0.7$, $\Psi(-10^6)=0.9$ and $\Phi'(-10^6)=\Psi'(-10^6)=0$
and the plots show perturbation evolution for $k=10^{-10}$. The plots
of the perturbation evolution in the Jordan frame are visibly
continuous and smooth. We will see later that if we try to get these
results from the Einstein frame calculation we will hit localized
singularities.

In the next section we will try to recast the problem in the Einstein
frame. A part of this calculation was incompletely done in an earlier
publication \cite{Paul:2014cxa} where the authors disregarded the velocity
potential of the fluid in the Einstein frame. In the present work we
complete and rectify the previous calculation.
\subsection{Einstein frame}

In Einstein frame the scalar perturbation in the longitudinal gauge is given as
\begin{eqnarray}
d\tilde{s}^{2}=\tilde{a}^{2}(\eta)\left[-(1+2\tilde{\Phi})d\eta^{2}
+(1-2\tilde{\Psi})\delta_{ij} dx^{i}dx^{j}\right]\,.
\label{delta_g_tilde}
\end{eqnarray}
In Einstein frame the gravitational theory is essentially GR and the energy-momentum tensor of the hydrodynamic matter and the scalar field are both diagonal, it is easy to check from the $ij-{\rm th}(i\neq j)$ component of the perturbed field equation that
\begin{equation}
\tilde{\Phi}=\tilde{\Psi}\,.
\end{equation}
This is different from the behavior of the above quantities in the
Jordan frame, because the energy-momentum tensor of the existing
matter component being diagonal does not necessarily implies the
equality of the two Bardeen potentials in a higher derivative gravity
theory. This is a good instance where switching to the Einstein frame
makes things easier. The Jordan frame Bardeen potentials can be
recovered from the Einstein frame Bardeen potential as follows
\cite{Mukhanov:1990me},
\begin{eqnarray}
\Phi= -\frac{2}{3}\left(\frac{F^{2}}{F'a}\right)\left[\left(\frac{a}{F}
\right)\tilde{\Phi}\right]'\,,\,\,\,\,
\Psi = \frac{2}{3}\left(\frac{1}{FF'a}\right)(aF^{2}\tilde{\Phi})'\,.
\label{Phi_Psi}
\end{eqnarray}
Later it will be shown that these connecting formulae breaks down in
the case of asymmetric bounce in the Jordan frame.

As the velocity potential cannot be in general neglected when treating
the scalar metric perturbations we rewrite the perturbation equations
using this new information.  Perturbed Einstein tensor components are:
\begin{eqnarray}
\delta G^0_0&=&\frac{2}{\tilde{a}^2}[-3\tilde{\mathcal{H}}(\tilde{\mathcal{H}}\tilde{\Phi}+\tilde{\Phi}^{\prime})+\nabla^{2}\tilde{\Phi} ]\,,\\
\delta G^0_i&=&\frac{2}{\tilde{a}^2}\left[\tilde{\Phi}^{\prime}+\tilde{\mathcal{H}}\tilde{\Phi} \right]_{,i}\,,\\
\delta G^i_j&=&-\frac{2}{\tilde{a}^2}[(2\tilde{\mathcal{H}}^\prime+\tilde{\mathcal{H}}^2)\tilde{\Phi}+3\tilde{\mathcal{H}}\tilde{\Phi}^\prime+\tilde{\Phi}^{\prime\prime}]\delta^i_j\,.
\end{eqnarray}
Here a partial derivative with spatial coordinates is specified with
the comma followed by a Latin alphabet. Perturbed energy-momentum tensor components for hydrodynamic matter are \cite{Mukhanov:1990me}:
\begin{eqnarray}
\delta \tilde{T}_0^{0}&=&\delta\tilde{\rho}\,,\\
\delta \tilde{T}_i^{0}&=&(\tilde{\rho}+\tilde{P})\tilde{a}^{-1}\delta \tilde{u}_i=-(\tilde{\rho}+\tilde{P})\tilde{a}^{-1}\partial_i \tilde{U}\,,
\label{tt0i}\\
\delta \tilde{T}_j^{i}&=&-\delta \tilde{P}\delta^i_j\,.
\end{eqnarray}
The quantity $\tilde{U}$ is the velocity potential given by $\delta
\tilde{u}_i=-\partial_i \tilde{U} +\tilde{v}_i$. Here $\tilde{v}_i$ is
is the pure vector part of the fluid velocity perturbation in the
Einstein. The perturbed scalar field energy momentum tensor components
are:
\begin{eqnarray}
\delta S_0^{0}&=&a^{-2}\left(-{\phi^{\prime}}^{2}\tilde{\Phi}+\phi^{\prime}\delta\phi^{{\prime}} 
+\tilde{a}^{2}V_{,\phi}\delta\phi\right)\,,\\
\delta S_i^{0}&=&a^{-2}\phi^{\prime}\delta\phi_{,i}\,,\\
\delta S_j^{i}&=&-a^{-2}\left(-\tilde{\Phi} {\phi^{\prime}}^{2}+\phi^{\prime} \delta\phi^{\prime} -\tilde{a}^{2} V_{,\phi}\delta\phi\right)\delta^i_j\,.
\end{eqnarray}
A derivative with respect to $\phi$ is specified by a comma followed by
$\phi$ in the subscript. Using the results we can write the perturbed
Einstein equations as:
\begin{eqnarray}
\label{rho}
-3\tilde{\mathcal{H}}(\tilde{\mathcal{H}}\tilde{\Phi}+\tilde{\Phi}^{\prime})+\nabla^{2}\tilde{\Phi}  &=& \frac{\kappa}{2} \left( \tilde{a}^{2}\delta \tilde{\rho} -{\phi^{\prime}}^{2}\tilde{\Phi} +\phi^{\prime}\delta\phi^{\prime} 
+\tilde{a}^{2}V_{,\phi} \delta\phi\right)\,,
\label{perte1}\\ 
(2\tilde{\mathcal{H}}^{\prime}+\tilde{\mathcal{H}}^{2}) \tilde{\Phi} +\tilde{\Phi}^{\prime \prime} +3\tilde{\mathcal{H}}\tilde{\Phi}^{\prime}  &=& \frac{\kappa}{2}\left( \tilde{a}^{2}\delta \tilde{P} -\tilde{\Phi} {\phi^{\prime}}^{2} + \phi^{\prime} \delta\phi^{\prime} -\tilde{a}^{2} V_{,\phi}\delta\phi\right)\,,\\
\frac{\kappa}{2}\left[\phi' \delta\phi-\tilde{a}(\tilde{\rho}+\tilde{p})  \tilde{U}\right]&=&\tilde{\Phi}^{\prime}+\tilde{\mathcal{H}}\tilde{\Phi}\,. 
\label{vpot}
\end{eqnarray}
Multiplying the first of the above set of equations by $c_s^2$ and
subtracting from the second one yields,
\begin{eqnarray}\label{b}
\tilde{\Phi}^{\prime \prime}- c_{s}^{2}\nabla^{2}\tilde{\Phi} +
\left( 2\tilde{\mathcal{H}}^{\prime} + 
\tilde{\mathcal{H}}^{2}
  \right)\tilde{\Phi} + 
3 \tilde{\mathcal{H}}\tilde{\Phi}^{\prime} + 
3 c_{s}^{2}\tilde{\mathcal{H}}\left(  \tilde{\mathcal{H}}\tilde{\Phi} 
+ \tilde{\Phi}^{\prime} \right) \nonumber\\
= -\frac{\kappa}{2} \tilde{\Phi}\phi^{\prime 2}
\left( 1- c_{s}^{2} \right)+\frac{\kappa}{2}\phi^{\prime}(1-c_{s}^{2})\delta
{\phi}^\prime - \frac{\kappa \tilde{a}^2}{2} V_{,\phi}
(1 + c_{s}^{2})\delta \phi\,.
\end{eqnarray}
This is one equation which gives the dynamics of the perturbed
potential $\tilde{\Phi}$ in the Einstein frame. The scalar field
perturbation is linked with the above dynamics. We require more
equations for uniquely solving the perturbation evolutions.

The other  equation comes from perturbing the Klein-Gordon equation as:
\begin{equation}
\tilde{\Box} \phi -V_{, \phi} + \frac{1}{\sqrt{-\tilde{g}}} \frac{\partial 
  \mathcal{L_M}}{\partial \phi}=0\,,
\label{kg1}
\end{equation}
where $\tilde{\Box} \equiv \tilde{D}^\mu \tilde{D}_\mu$,
$\tilde{D}_\mu$ being the covariant derivative in the Einstein
frame. Here $\mathcal{L_M}$ specifies the Lagrangian of the
hydrodynamic fluid. Using the following fact  we can write down the
last term on the left hand side of the above equation,
$$\frac{\partial \mathcal{L_M}}{\partial \phi} =  \frac{\partial 
\mathcal{L_M}}{\partial g^{\mu \nu}} \frac{\partial
  g^{\mu \nu}}{\partial \phi} = \frac{1}{F(\phi)}\frac{\partial 
\mathcal{L_M}}{\partial \tilde{g}^{\mu \nu}} \frac{\partial
 (F(\phi)\tilde{g}^{\mu \nu})}{\partial \phi}\,.$$
Using the standard definition of the matter energy momentum tensor 
$$\tilde{T}_{\mu \nu} = - \frac{2}{\sqrt{-\tilde{g}}} \frac{\partial 
  \mathcal{L_M}}{\partial \tilde{g}^{\mu \nu}}\,,$$
we can write,
$$\frac{\partial \mathcal{L_M}}{\partial \phi} = -\sqrt{-\tilde{g}}
\frac{F_{,\phi}}{2F}\tilde{g}^{\mu \nu} \tilde{T}_{\mu \nu}=-\sqrt{-\tilde{g}}
\frac{F_{,\phi}}{2F} \tilde{T}\,,$$
where $\tilde{T}$ is the trace of the energy-momentum tensor in the Einstein frame.
In metric $f(R)$ cosmology it is always, 
$$\frac{F_{,\phi}}{2F}=\sqrt{\frac{\kappa}{6}}\,,$$
and consequently Eq.~(\ref{kg1}) becomes
\begin{equation}
\tilde{D}^\mu \tilde{D}_\mu \phi -V_{, \phi}  -\sqrt{\frac{\kappa}{6}}\tilde{T} =0\,.
\label{kg2}
\end{equation}
Perturbing the terms in the Klein-Gordon equation, without the matter
coupling term, one gets
\begin{eqnarray}
\delta(\tilde{D}^\mu \tilde{D}_\mu \phi - V_{, \phi})
&=&-\tilde{a}^{-2}\left[\delta \phi^{\prime \prime} +
  2\tilde{\mathcal{H}}\delta \phi^\prime - \nabla^2 \delta \phi -
  4\phi^\prime \tilde{\Phi}^\prime + \tilde{a}^2 V_{, \phi \phi}
  \delta \phi +2 \tilde{a}^2 V_{, \phi}
  \tilde{\Phi}\right.\nonumber\\ &&\left.- 2 \tilde{a}^2
  \sqrt{\frac{\kappa}{6}} (1-3c_s^2) \tilde{\Phi} \tilde{\rho}
  \right]\,,
\label{dkg}
\end{eqnarray}  
where we have used the background Klein-Gordon equation for the scalar
field in the Einstein frame. The perturbation of the
matter coupling term gives
\begin{eqnarray}
\delta \tilde{T} = \delta \tilde{g}^{\mu \nu} \tilde{T}_{\mu \nu}
+ \tilde{g}^{\mu \nu} \delta \tilde{T}_{\mu \nu}=-(1-3c_s^2) \delta \tilde{\rho}\,.
\label{dt}
\end{eqnarray}
Combining the results the perturbed Klein-Gordon equation gives,
\begin{eqnarray}
\delta \phi^{\prime \prime} &+& 2\tilde{\mathcal{H}}\delta \phi^\prime - \nabla^2 \delta \phi
-4\phi^\prime \tilde{\Phi}^\prime + \tilde{a}^2 V_{, \phi \phi} \delta \phi
+2 \tilde{a}^2 V_{, \phi} \tilde{\Phi} -2\tilde{a}^2 \sqrt{\frac{\kappa}{6}}
 (1-3c_s^2) \tilde{\Phi} \tilde{\rho}\nonumber\\
&-&\tilde{a}^2 \sqrt{\frac{\kappa}{6}} (1-3c_s^2) \delta \tilde{\rho}=0\,.
\label{dkgc}
\end{eqnarray}
\begin{figure}[t!]
\begin{minipage}[b]{0.5\linewidth}
\centering
\includegraphics[scale=.5]{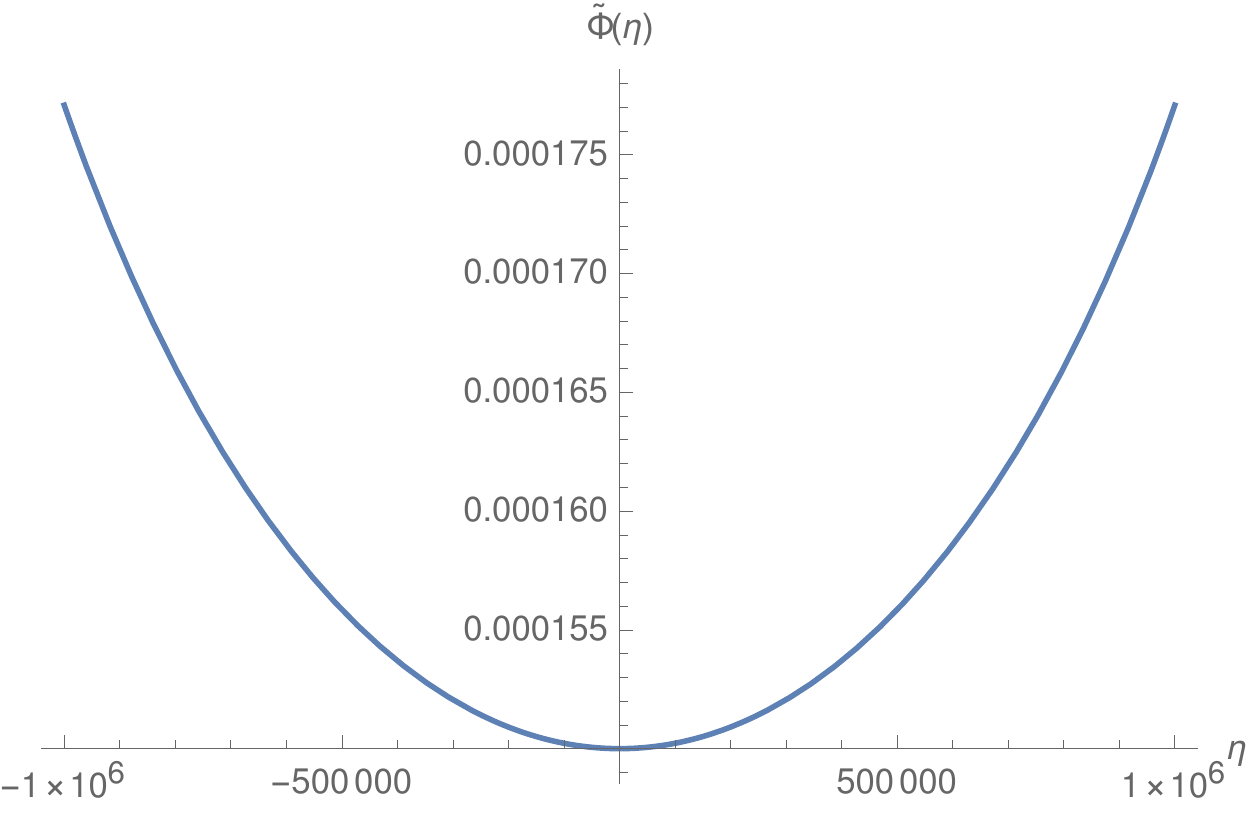}
\caption{Evolution of $\tilde{\Phi}$ with $\eta$ through a symmetric bounce in the background.} 
\label{an1}
\end{minipage}
\hspace{0.2cm}
\begin{minipage}[b]{0.5\linewidth}
\centering
\includegraphics[scale=.5]{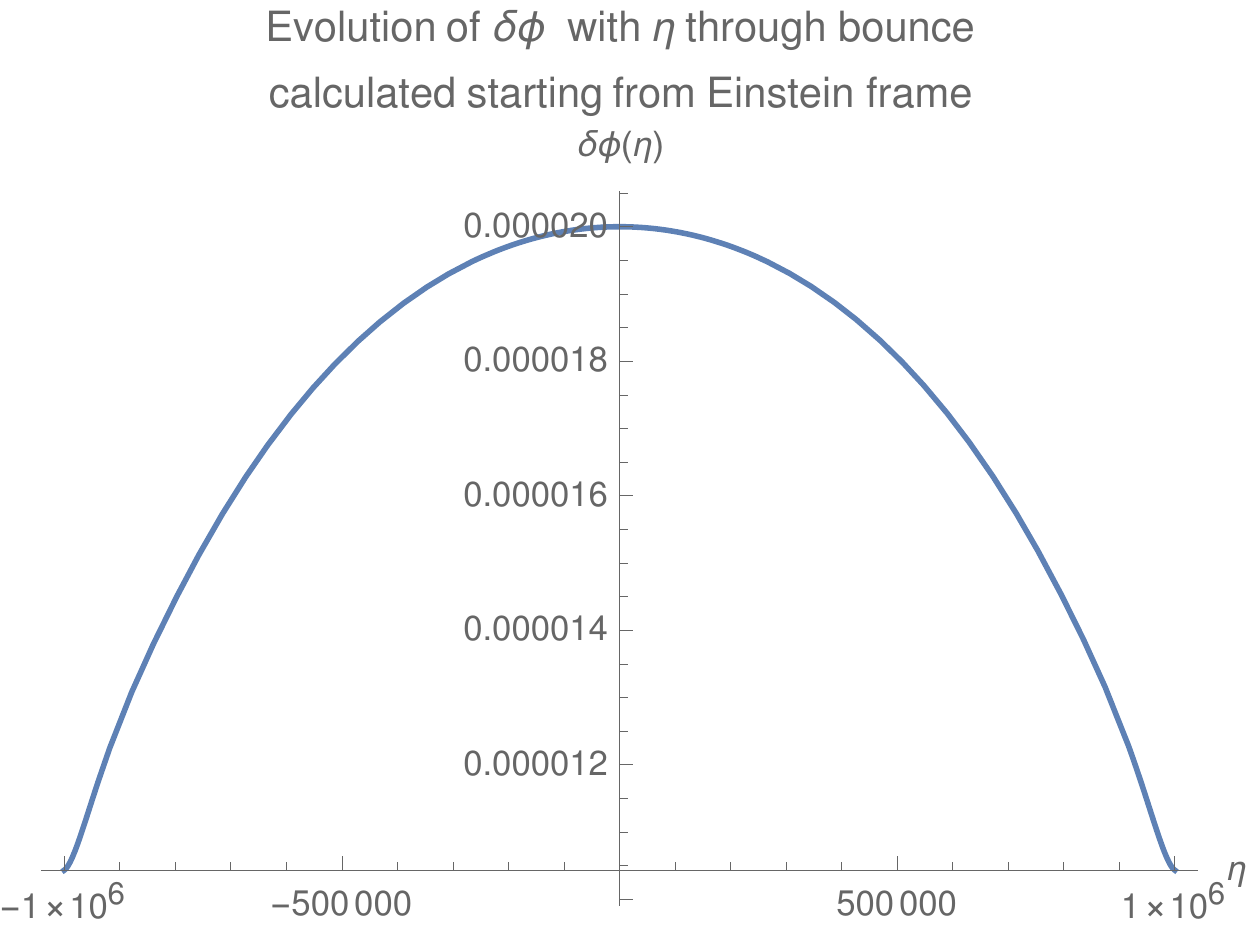}
\caption{Evolution of $\delta\phi$ with $\eta$ through a symmetric bounce in the background.}
\label{endens1}
\end{minipage}
\end{figure}
Using the expression of $\delta\tilde{\rho}$ from Eq.~(\ref{perte1}) we can write the above
equation as
\begin{multline}\label{perkg}
\delta \phi^{\prime \prime} -\nabla^2 \phi + 2 \tilde{\Phi}
\tilde{a}^2 V_{, \phi} + \sqrt{\frac{\kappa}{6}} (1- 3c_{s}^{2}) \Big(
-\phi^{\prime 2} -2 \tilde{a}^2 \tilde{\rho} + \frac{6
  \tilde{\mathcal{H}}^2}{\kappa} \Big)\tilde{\Phi} \\
-2 \sqrt{\frac{\kappa}{6}} (1- 3c_{s}^{2})\frac{\nabla^2
\tilde{\Phi}}{\kappa}-\left[4 \phi^{\prime} - \sqrt{\frac{\kappa}{6}} (1-
  3c_{s}^{2}) \frac{6 \tilde{\mathcal{H}}^2}{\kappa}\right] \tilde{\Phi}^{\prime}\\
+\left[V_{, \phi \phi} + \sqrt{\frac{\kappa}{6}} (1- 3c_{s}^{2})V_{, \phi}\right]
\tilde{a}^2 \delta \phi+ \left[2 \tilde{\mathcal{H}} + \sqrt{\frac{\kappa}{6}} (1-
3c_{s}^{2})\phi^{\prime}\right] \delta \phi^{\prime} =0\,.
\end{multline}
One can now solve solve Eq.~(\ref{perkg}) and Eq.~(\ref{b})
simultaneously and obtain the evolution of $\tilde{\Phi}$ and
$\delta\phi$.  These are the general results related to evolution of
scalar metric perturbations in the longitudinal gauge worked out in
the the Einstein frame which are appropriate for early universe
cosmological processes as bounce. Previous authors have worked the
Einstein frame perturbation equations in the synchronous gauge
\cite{Bean:2006up} to study the problem of structure formation in the
late time universe. The results obtained in the cited work cannot in
general be applied to study the problem of scalar metric perturbation
evolution through a non-singular bounce and to our knowledge the
appropriate longitudinal gauge results for scalar perturbation growth
which we present in this paper are reported for the first time.  Next
we will apply the formalism in the case of a background exponential
bounce. For the particular $f(R)$ model as chosen in
Eq.~(\ref{frform}) the scalar field potential $V(\phi)$ is given by:
\begin{equation}
V(\phi)= \frac{1}{2 \kappa \alpha} \Big(\sqrt{\frac{2 \kappa}{3}} \phi- 1 \Big) e^{- \sqrt{\frac{2 \kappa}{3}}\phi}\,,
\label{potf}  
\end{equation}
where $\alpha= 10^{12}$. The above equations are important results
reported for the first time in this paper.
\subsubsection{Using the Einstein frame to model symmetric bounces in the Jordan frame}
\label{sse}
\begin{figure}[t!]
\begin{minipage}[b]{0.5\linewidth}
\centering
\includegraphics[scale=.5]{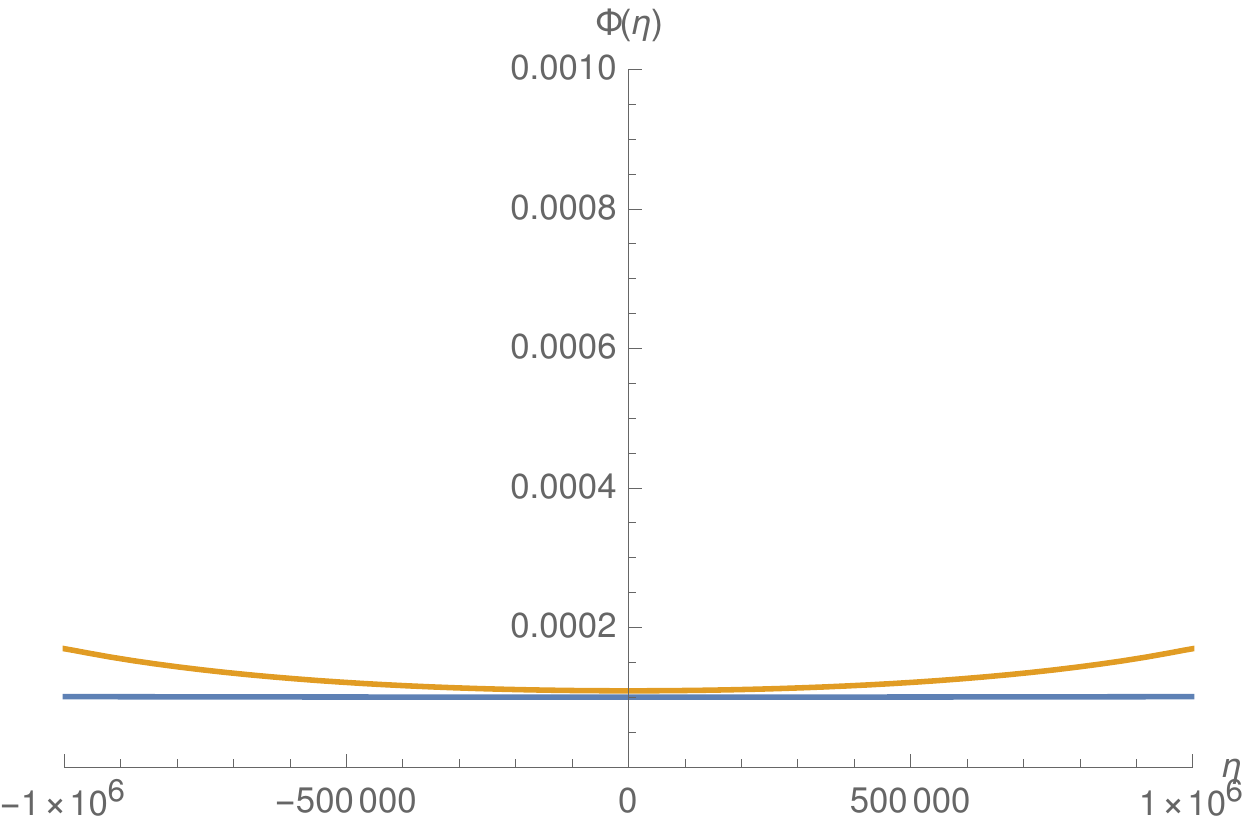}
\caption{Evolution of $\Phi$ with $\eta$ through symmetric bounce. The
  blue curve represents Jordan frame result and the orange curve
  represents the result obtained from the Einstein frame. The details
  regarding the plots are given in text.} 
\label{pn2}
\end{minipage}
\hspace{0.2cm}
\begin{minipage}[b]{0.5\linewidth}
\centering
\includegraphics[scale=.5]{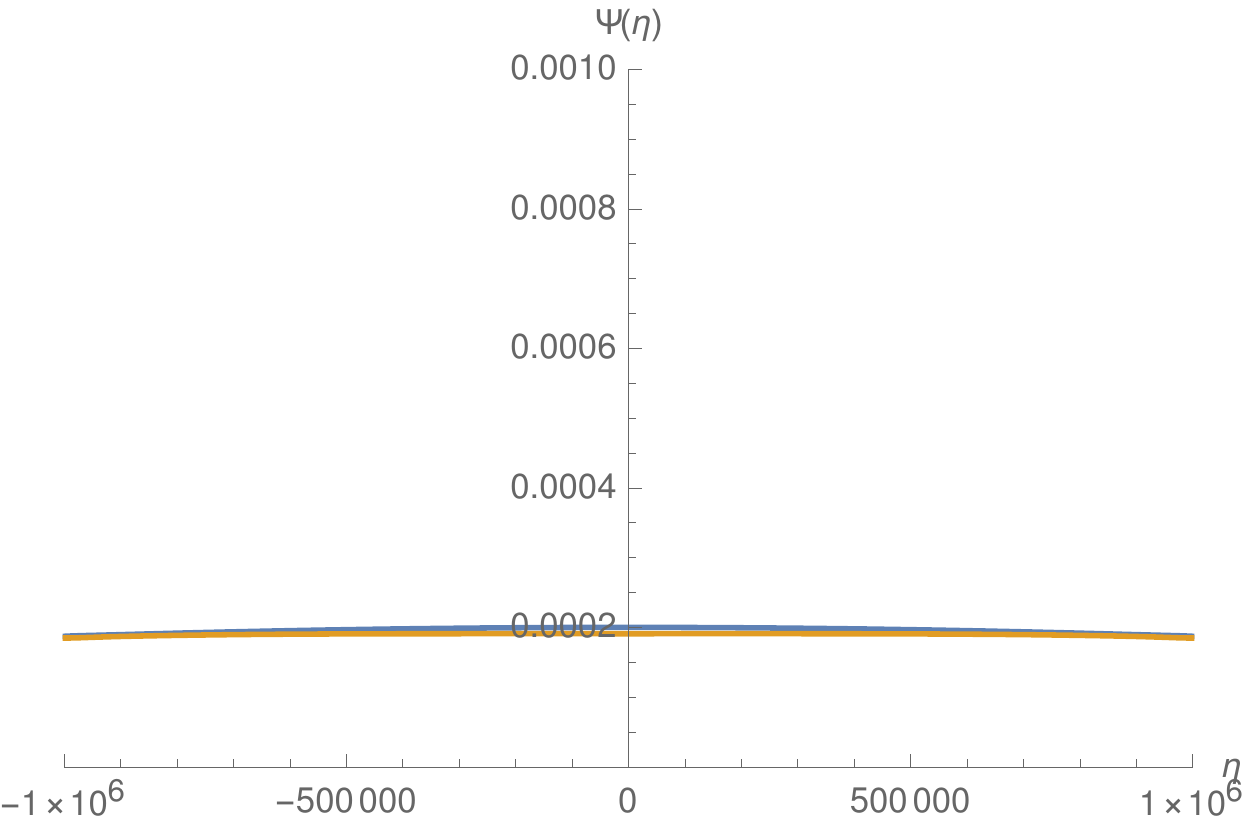}
\caption{Evolution of $\Psi$ with $\eta$ through symmetric bounce. The
  blue curve represents Jordan frame result and the orange curve
  represents the result obtained from the Einstein frame. The details
  regarding the plots are given in text.}
\label{pn3}
\end{minipage}
\end{figure}

We can choose the conditions in the Einstein frame variables in such a
way so that the dynamics produces a symmetric bounce in the Jordan
frame. We will choose the conditions in the Einstein frame so that we
can reproduce the results of the symmetric bounce in the Jordan frame
as presented in the previous subsection, modulo some small numerical
error which arises in the program to convert the result from Einstein
frame to Jordan frame. In the Einstein frame we choose $\phi(0)=0.1$,
$\phi^\prime(0)=0$ for the symmetric bounce background. For the
perturbations we use $\tilde{\Phi}(0)=0.00015$,
$\tilde{\Phi}^\prime(0)=0$, $\delta \phi(0)=.00002$ and $\delta
\phi^\prime(0)=0$. The above set of values are not chosen randomly,
they are chosen in such a way such that they reproduce the analogous
conditions at $\eta=0$ imposed in Jordan frame perturbation
calculations for the symmetric bounce, presented in the previous
subsection.  The background cosmological evolution is guided by the
equations given at the last part of subsection \ref{jefr}. The
background evolution was specified in terms of coordinate time in the
Einstein frame and we use the same equations to produce the background
dynamics. After the background dynamics is done in coordinate time we
map the results to conformal time so that the background calculation
matches with perturbation dynamics results presented in this
subsection. Fig.~\ref{an1} and Fig.~\ref{endens1} shows the evolution
of $\tilde{\Phi}$ and $\delta \phi$ in the Einstein frame for a
symmetric cosmological bounce in the background Jordan frame. One can
now use the relations given in Eq.~(\ref{Phi_Psi}) to convert the
perturbations from the Einstein frame to the Jordan frame. We expect
that the results so obtained will closely match with the results
obtained by the perturbation dynamics calculations done in the Jordan
frame, plotted in blue in  Fig.~\ref{pn2} and Fig.~\ref{pn3}.
The present results are plotted in orange, in Fig.~\ref{pn2} and
Fig.~\ref{pn3}. We see that our results do match to a great extent,
the slight mismatch results from the numerical techniques utilized in
calculating the background and perturbation dynamics in the two
separate frames.
\subsubsection{Using the Einstein frame to model asymmetric bounces in the Jordan frame}
\begin{figure}[t!]
\begin{minipage}[b]{0.5\linewidth}
\centering
\includegraphics[scale=.5]{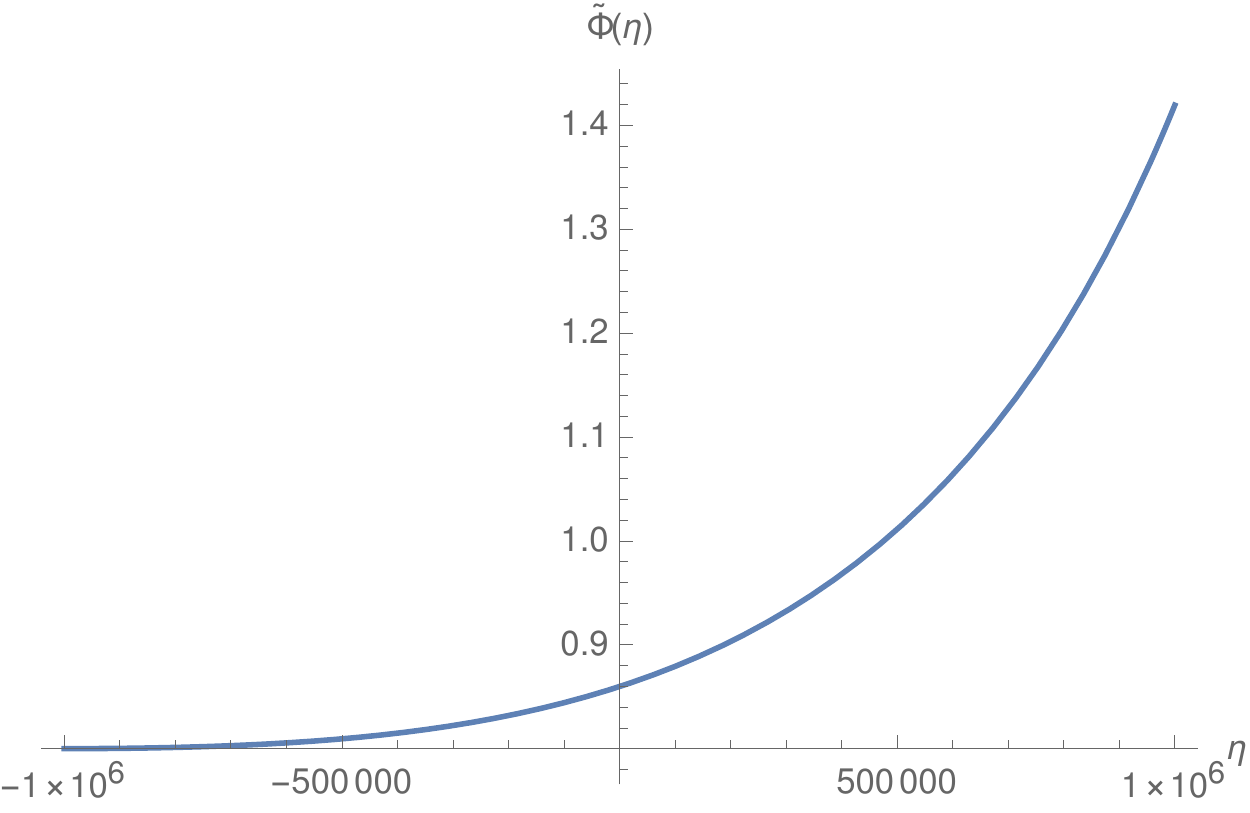}
\caption{Evolution of $\tilde{\Phi}$ with $\eta$, in the case of an asymmetric
  bounce in Jordan frame, calculated directly in Einstein frame.} 
\label{panb}
\end{minipage}
\hspace{0.2cm}
\begin{minipage}[b]{0.5\linewidth}
\centering
\includegraphics[scale=.5]{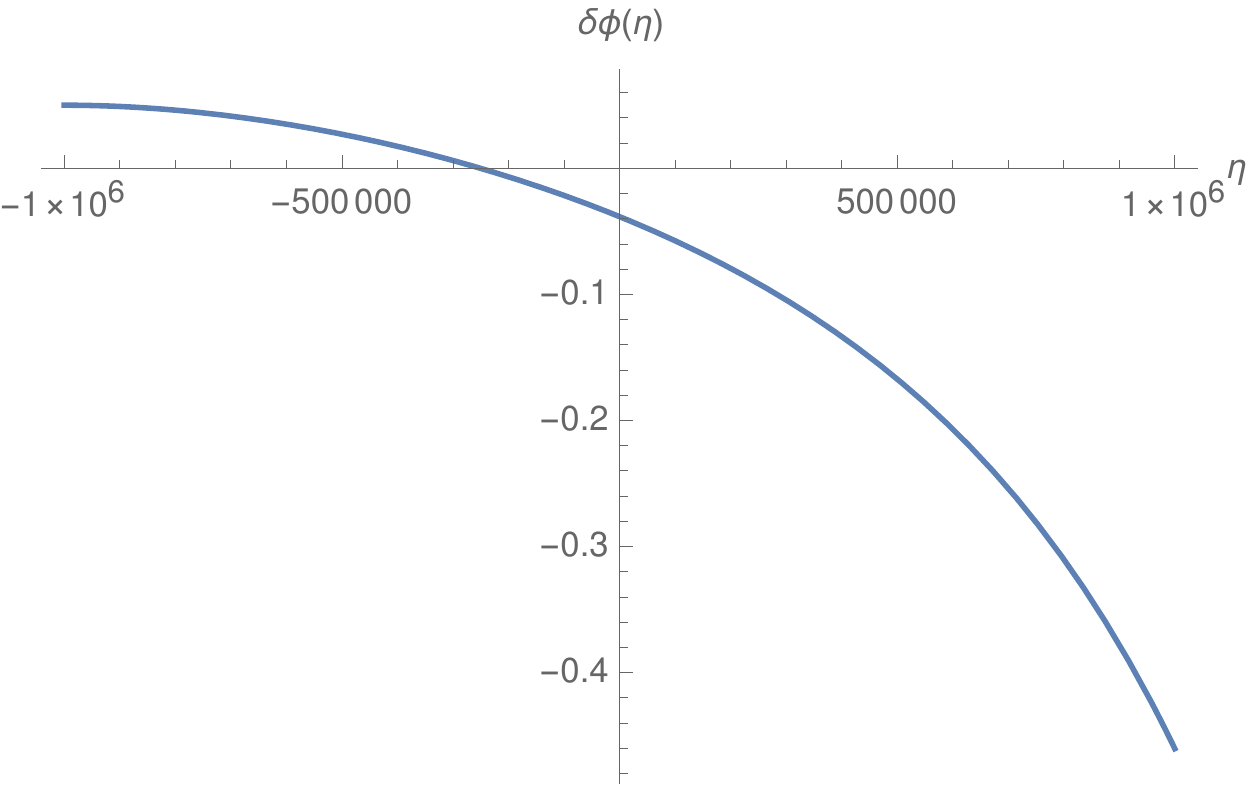}
\caption{Evolution of $\delta{\phi}$ with $\eta$, in the case of an
  asymmetric bounce in Jordan frame, calculated directly in Einstein frame.}
\label{pana}
\end{minipage}
\end{figure}

\begin{figure}[t!]
\begin{minipage}[b]{0.5\linewidth}
\centering
\includegraphics[scale=.5]{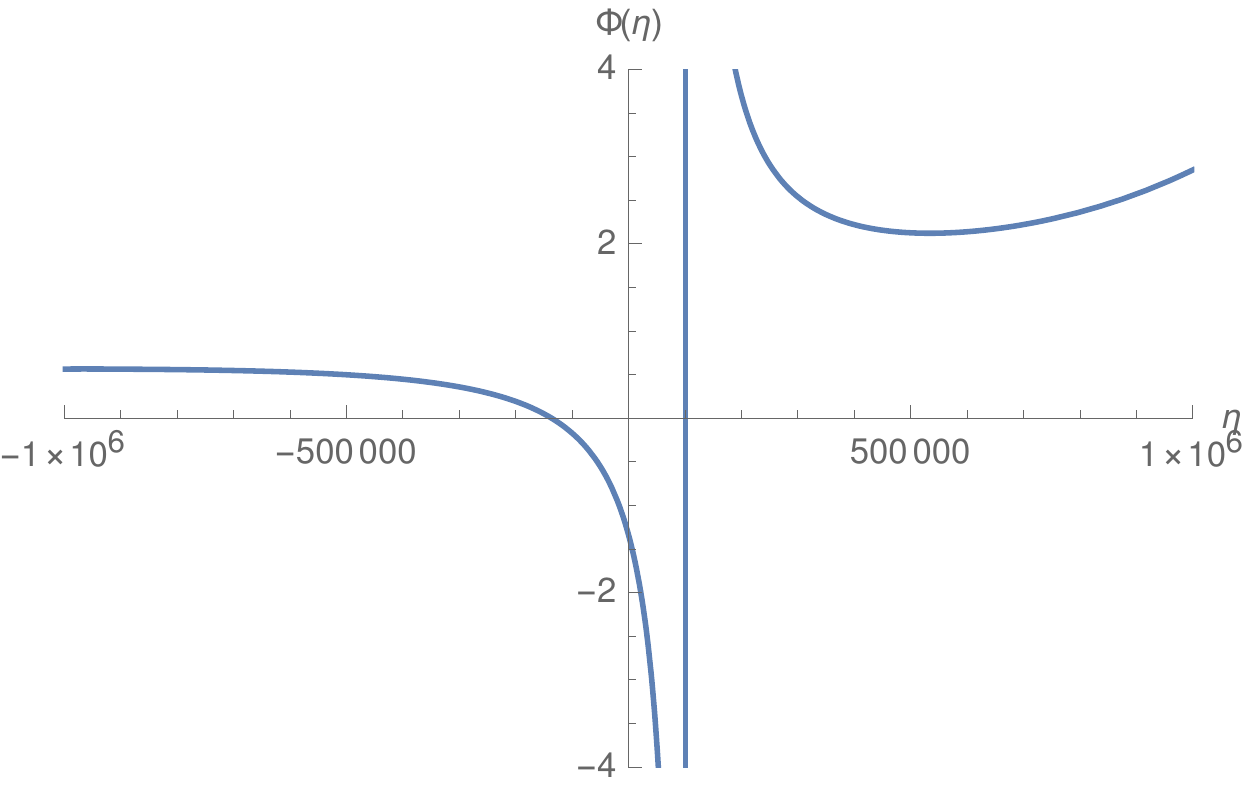}
\caption{Evolution of $\Phi$ with $\eta$ through an asymmetric bounce in the Jordan frame obtained from the results in the Einstein frame.} 
\label{pnc}
\end{minipage}
\hspace{0.2cm}
\begin{minipage}[b]{0.5\linewidth}
\centering
\includegraphics[scale=.5]{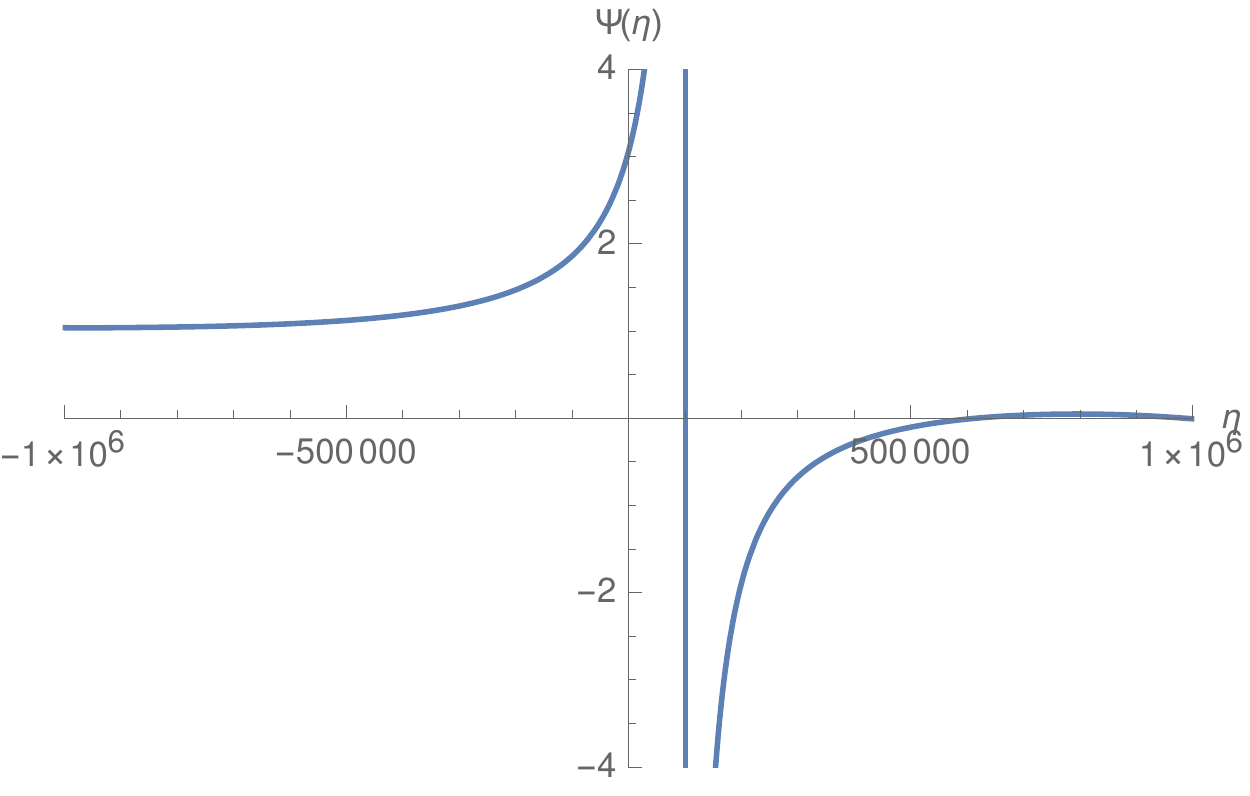}
\caption{Evolution of  $\Psi$  with $\eta$ through an asymmetric bounce in the Jordan frame obtained from the results in the Einstein frame.} 
\label{pnew}
\end{minipage}
\end{figure}

If we consider an asymmetric bounce, we see that the nature of the
scalar perturbations calculated from the two frames do not match with
each other. This fact was partially noted in Ref.~\cite{Paul:2014cxa}, in the
present article we fully specify the gravity of the situation. At
first we point out the most important difference between perturbation
dynamics for the symmetric and asymmetric bounce case. In the case of
an asymmetric bounce in the Jordan frame we can smoothly plot the
perturbations $\Phi$, $\Psi$ in the Jordan frame as shown in
Fig.~\ref{pb} and Fig.~\ref{pa} or $\tilde{\Phi}$, $\delta \phi$ in
the Einstein frame as shown in Fig.~\ref{panb} and Fig.~\ref{pana}.
In this paper all the calculations of the perturbation dynamics is
done for the Fourier mode $k=10^{-10}$. The asymmetric bounce in the
Jordan frame can be obtained from an Einstein frame cosmology where
$\phi(0)=0.1$ and $\phi^\prime(0)=10^{-8}$.  These values corresponds
to the values for $\cal{H}^\prime$ and $\cal{H}^{\prime \prime}$
applied in the Jordan frame to produce an asymmetric bounce.  To plot
the perturbations in the Einstein frame we have used the conditions
$\tilde{\Phi}(-10^6)=0.8, \tilde{\Phi}'(-10^6)= 0$ and $\delta
\phi(-10^6)= 0.05, \delta \phi'(-10^6)=0$. These values match with the
analogous conditions used to plot the scalar perturbations in the case
of an asymmetric bounce in the Jordan frame as shown in Fig.~\ref{pb}
and Fig.~\ref{pa}. In the Einstein frame $\tilde{\Phi}$ becomes
marginally non-perturbative after for $\eta>0$ but the actual Jordan
frame scalar perturbations remain perturbative within the time period
of our interest. The marginal non-perturbative behavior in the
Einstein frame do not posit any cosmological problem as far as the
bounce in the Jordan frame is considered. The interesting feature of
the asymmetric bounce appears when one tries to calculate the Jordan
frame perturbation evolution using the Einstein frame results via the
use of the relations in Eq.~(\ref{Phi_Psi}).
Using the relations in Eq.~(\ref{Phi_Psi}) one obtains the Jordan
frame results shown in Fig.~\ref{pnc} and Fig.~\ref{pnew}. The results
do not match with the expected result as obtained in Fig.~\ref{pb} and
Fig.~\ref{pa}. The transformation from the Einstein frame to the
Jordan frame produces unavoidable singularities. The singularities arise
in the case of an asymmetric bounce due to the reason that
$F^\prime=0$ for some $\eta \ne 0$ in the time period of our interest,
making the relations in Eq.~(\ref{Phi_Psi}) singular. For symmetric
bounces one has $F^\prime(0)=a^\prime(0)=\tilde{\Phi}(0)=0$ and the
singularity disappears in the $\eta \to 0$ limit in
Eq.~(\ref{Phi_Psi}) as both the numerator and denominator tends to
vanish at the same time instant. The point was partially discussed in
\cite{Paul:2014cxa}. In the present case the singularities lie near
$\eta=0$ and so the Jordan frame potentials blow up near $\eta=0$ when
we apply the transformations in Eq.~(\ref{Phi_Psi}) to the Einstein
frame results. If the conditions used to plot the perturbations were
applied at $\eta=0$ (in both the frames) then the blowing up of the
potentials near $\eta=0$ distorts the values of $\Phi$ and $\Psi$
obtained far from $\eta=0$. In this case it is better to use initial
conditions at $\eta=-10^6$, which is much further from the
singularities encountered in the transformations. Before we finish
this discussion we must point out that the singularities shown in the
perturbation evolutions in the asymmetric case are not real
singularities but an artefact of using a conformal frame which is not
suitable to tackle asymmetric bounces. As a consequence our results
show that the Einstein frame can be used to calculate most of the
properties of a symmetric cosmological bounce in the Jordan frame
including the scalar perturbation evolution. In this case the Einstein
frame actually serves as an auxiliary conformal frame where the
calculations can be done and the results can be converted back to the
Jordan frame. On the other hand for an asymmetric bounce the Einstein
frame can act as a true auxiliary frame for the background evolution
but fails to reproduce the scalar perturbation dynamics in the Jordan
frame. This fact is of paramount importance showing that the Jordan
frame is the natural choice for scalar metric perturbation dynamics.
\subsubsection{Evolution of velocity potential in both the frames}

\begin{figure}[t!]
\centering
\includegraphics[scale=.5]{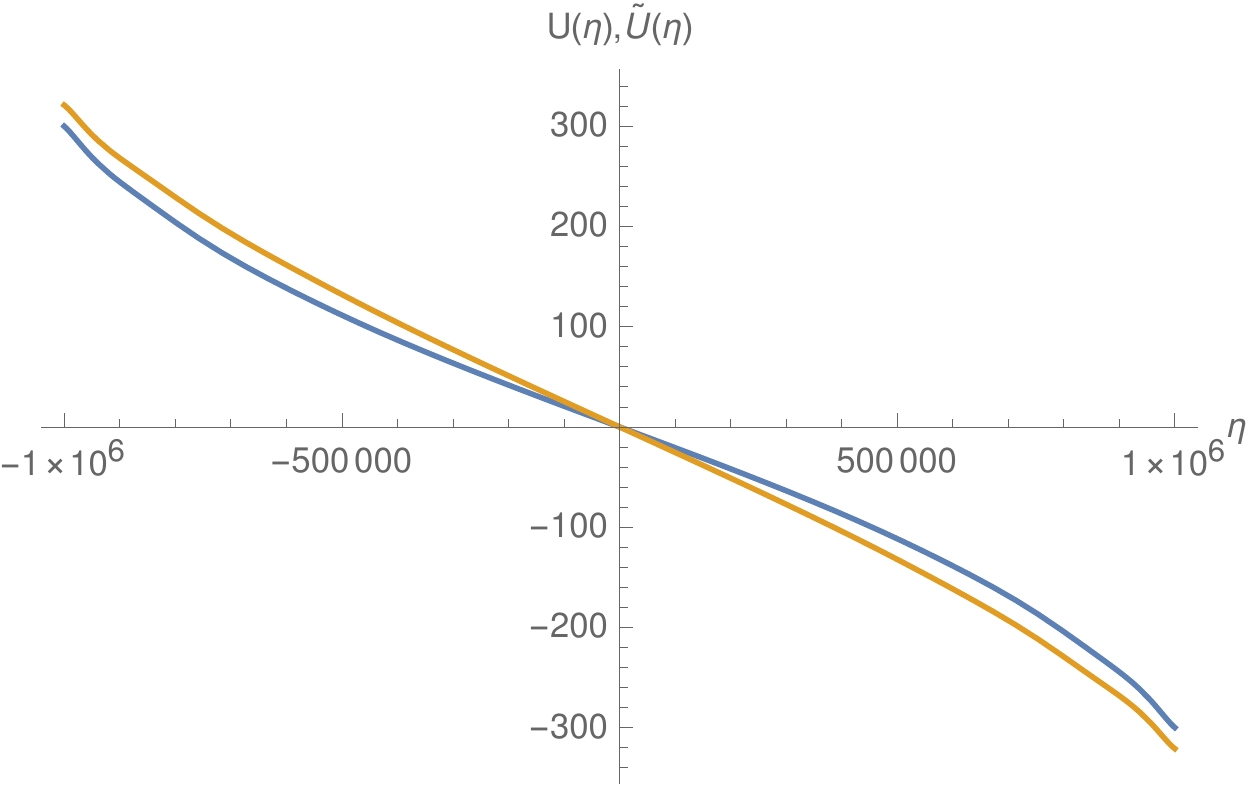}
\caption{Evolution of fluid velocity potential $U$ and $\tilde{U}$ for
  a symmetric bounce in the Jordan frame. The blue curve represents
  the Jordan frame result and the orange curve represents the result
  obtained from the Einstein frame.}
\label{utf}
\end{figure}

We omitted the equation involving fluid velocity potential in the
Jordan frame as the potential can be calculated from the Einstein
frame itself. After showing that the scalar perturbations can be
correctly calculated from the Einstein frame we directly use the
Einstein frame to predict the nature of the fluid velocity potential
in the Jordan frame.  In the Einstein frame Eq.~(\ref{vpot}) can be
used to predict the evolution of the velocity potential
$\tilde{U}$. Once the evolution $\tilde{U}$ is known one can convert
the result to the Jordan frame to opine on the behavior of $U$.  We
have $\tilde{T}^\mu_\nu=T^\mu_\nu/F^2$ and consequently
$$\delta \tilde{T}^0_i= \frac{\delta T^0_i}{F^2} -2 \frac{\delta T^0_i}{F^3}\,.$$
As $ T^0_i=0$ and the form of $\delta \tilde{T}^0_i$ (or $\delta T^0_i$) can be obtained from
the form of Eq.~(\ref{tt0i}) we can write
$$(\tilde{\rho} + \tilde{P})\tilde{a}^{-1} \partial_i \tilde{U}
=\frac{(\rho + P)a^{-1} \partial_i U}{F^2}\,,$$ which yields
\begin{eqnarray}
\tilde{U} = \sqrt{F} U\,.
\label{uut}
\end{eqnarray}
The above equation shows how the velocity potentials in the two
conformal frames are related to each other. The plots of the velocity
potentials, when $k=10^{-10}$, for a symmetric bounce in the
background is shown in Fig.~\ref{utf}. The blue curve represents the
Jordan frame result and the orange one represents the result obtained
from the Einstein frame. The results reasonably match with each other.
\section{Vector perturbations in $f(R)$ cosmology}
\label{vp}
In this section we will specialize on vector metric perturbations and
try to see how these perturbations evolve in $f(R)$ cosmology. In GR
based bouncing models of cosmology the metric vector perturbation is
bound to grow in the contracting phase of the universe \cite{Battefeld:2004cd}. But such a
behavior is not in general true in $f(R)$ bouncing models as we will
see below. The behavior of vector perturbations in $f(R)$ cosmology
is modulated by the behavior of $F(R)$ which can keep the vector
perturbations under tight control.

In the Jordan frame the metric is written as
\[
\renewcommand\arraystretch{1.3}
 g_{\mu\nu}=a^2(\eta)
\mleft[
\begin{array}{c|c}
  -1 & S_{i} \\
  \hline
  S_{i} &  \delta_{ij}-Q_{i,j}-Q_{j,i}
\end{array}
\mright]\,,
\]
and
\[
\renewcommand\arraystretch{1.3}
 g^{\mu\nu}=\frac{1}{a^2(\eta)}
\mleft[
\begin{array}{c|c}
  -1 & S^{i} \\
  \hline
  S^{i} &  \delta^{ij}+Q^{i,j}+Q^{j,i}
\end{array}
\mright]\,,
\]
where $S^i$ and $Q^i$ ate 3-vectors satisfying the constraint
$S^i_{\,\,,i}=Q^i_{\,\,,i}=0$.  The metric in the Einstein frame is
given in an identical way except that $\tilde{a}$, $\tilde{S_i}$ and
$\tilde{Q_{i,j}}$ appear instead of $a$, $S_i$ and $Q_{i,j}$. More
over the metric perturbations remain the same in both the conformal
frames. In the Jordan frame the relevant quantities calculated from
the metric given above are:
\begin{align}
R_{00}= -3 \mathcal{H}^\prime\,,\,\,\,\,\,
R_{0i}=(\mathcal{H}^\prime + 2 \mathcal{H}^2) S_{i} -\frac{1}{2} \nabla S_i\,,
\end{align}
where $\nabla\equiv \partial_j \partial^j$, and
\begin{multline}
  R_{ij}= (\mathcal{H}^\prime + 2 \mathcal{H}^2) \delta_{ij}-\frac{1}{2}(S^\prime_{i,j}+
  S^\prime_{j,i})- \mathcal{H}(S_{i,j}+ S_{j,i})\\
  -(\mathcal{H}^\prime + 2 \mathcal{H}^2)(Q_{i,j}+ Q_{j,i})-\mathcal {H} (Q^\prime_{i,j} +
  Q^\prime_{j,i})- \frac{1}{2} (Q^{\prime\prime}_{i,j} + Q^{\prime\prime}_{j,i})\,.
\end{multline}
The Ricci tensor remains unchanged:
\begin{equation}
R=  6\frac{(\mathcal{\dot{H}}+\mathcal{H}^2)}{a^2}\,,
\end{equation}
as $\delta R=0$. The perturbed Einstein tensor components are
\begin{equation}
\delta G^{0}_{0}=0 \,,\,\,\,\,
\delta G^{0}_{i}= \frac{1}{2a^2} \nabla S_i\,,
\label{einv1}
\end{equation}
and 
\begin{eqnarray}
\delta G^{i}_{j} = -\frac{1}{2a^2}(S^{\prime i}_{\,\,\,,j}+ S^{\prime \,\,i}_{j,})- \frac{\mathcal{H}(S^{i}_{\,\,,j}+ S^{\,\,\,i}_{j,})}{a^2} -\frac{1}{2a^2}(Q^{\prime \prime i}_{\,\,\,\,\,\,,j}+ Q^{\prime \prime \,\,i}_{j,})- \frac{\mathcal{H}(Q^{\prime i}_{\,\,\,,j}+ Q^{\prime \,\,i}_{j,})}{a^2}\,,
\end{eqnarray}
where in our convention, $G_{\mu \nu} = R_{\mu \nu} - (1/2) g_{\mu
  \nu} R$, and, $\delta G^{\mu}_{\nu}= \delta (g^{\mu \alpha}
G_{\alpha \nu}) = \delta g^{\mu \alpha} G_{\alpha \nu}+ g^{\mu \alpha}
\delta G_{\alpha \nu}\,.$ The field equation in $f(R)$ theory is
\cite{Sotiriou:2008rp}:
\begin{equation}
G_{\mu \nu}= \frac{\kappa T_{\mu \nu}}{F(R)} + \frac{g_{\mu
    \nu}[f(R)-RF(R)]}{2F(R)}+\frac{D_{\mu} D_{\nu} F(R)- g_{\mu
    \nu}\Box F(R)}{F(R)}\,,
\end{equation}
where $D_\mu A^\nu \equiv \partial_\mu A^\nu + \Gamma^\nu_{\mu
  \lambda} A^\lambda$ represents a covariant derivative of a
contravariant 4-vector $A^\mu$ and $\Box \equiv g^{\mu \nu} D_\mu
D_\nu$.  Perturbing the above field equation one obtains
\begin{eqnarray} \label{2nd}
\delta G^{\mu}_{ \nu} &=& \frac{\kappa \delta T^{\mu}_{\nu}}{F} -
\frac{\kappa T^{\mu}_{ \nu }
F_R \delta R}{F^2} - \frac{\delta ^{\mu }_{\nu} f F_R \delta R}{2 F^2}-
\frac{\left[D^{\mu}D_{\nu} F- \delta ^{\mu }_{\nu} \Box F\right]F_R \delta R}
{F^2(R)} \nonumber\\
&+& \frac{1}{F} \left[\delta g^{\mu \alpha}( {\partial_{\alpha} \partial_{\nu} F -
\Gamma^{\lambda}_{\alpha \nu} \partial_{\lambda} F} )
+g^{\mu \alpha }\left({ \partial_{\alpha} \partial_{\nu}(F_R \delta R) -
\delta \Gamma^{\lambda}_{\alpha \nu} \partial_{\lambda} F -
\Gamma^{\lambda}_{\alpha \nu}\partial_{\lambda}(F_R \delta R)}\right)
\right.\nonumber\\
&-& \delta ^{\mu }_{\nu} ( \partial^{\lambda} \partial_{\lambda}(F_R \delta R) + \delta \Gamma^{\lambda}_{\rho \lambda} g^{\rho \kappa }(\partial_{\kappa} F)+ \Gamma^{\lambda}_{\rho \lambda} \delta g^{\rho \kappa }(\partial_{\kappa} F) + \Gamma^{\lambda}_{\rho \lambda} g^{\rho \kappa } \partial_{\kappa}(F_R \delta R))]\,.
\end{eqnarray}
For further progress we require the perturbed fluid energy-momentum
tensor in the Jordan frame, whose background value is specified in
Eq.~(\ref{tmunu}). In our convention we specify $u_\mu = -a(1,v_i)$ where the scale-factor is expressed in conformal time and $\delta u_i \equiv -a v_i$. In such a case one can easily see that $u^\mu = a^{-1}(1, -S^i-v^i)$. By perturbing Eq.~(\ref{tmunu}) one gets the non-zero components of $T^\mu_\nu$ in the absence of any anisotropic stress:
\begin{equation}
\delta T^{0}_{i} =  - (\rho + P)v_i \,,\,\,\,\,
\delta T^{i}_{0} =   (\rho + P)(S^i + v^i)\,,
\label{texp}
\end{equation}
where $\rho$ and $P$ are the background values of the thermodynamic
variables. From the above set of equations one can write the dynamical
equations for the perturbations as
\begin{align}
\frac{1}{2a^2} \nabla P^{i}&= -\frac{\kappa  }{F}  (\rho + P)v^{i}\,,
\label{fv1}\\
\frac{1}{2a^2}\partial_{\eta}\left[a^2(P^{i}_{\,\,,j}+ P^{\,\,\,i}_{j,})\right]&= -\frac{F^\prime}{2F} (P^{i}_{\,\,,j}+ P^{\,\,\,i}_{j,})\,,
\label{fv2}
\end{align}
where\\
\begin{equation}\label{118}
S^{i}+ Q^{\prime i} \equiv P^i
\end{equation}
is a gauge-invariant quantity. Henceforth we will work in the Newtonian gauge where $Q^i=0$.
One can easily check that in the limit when $f(R)=R$ the above equations become identical to the equations for vector perturbations obtained in \cite{Battefeld:2004cd}.
\begin{figure}[t!]
\begin{minipage}[b]{0.5\linewidth}
\centering
\includegraphics[scale=.6]{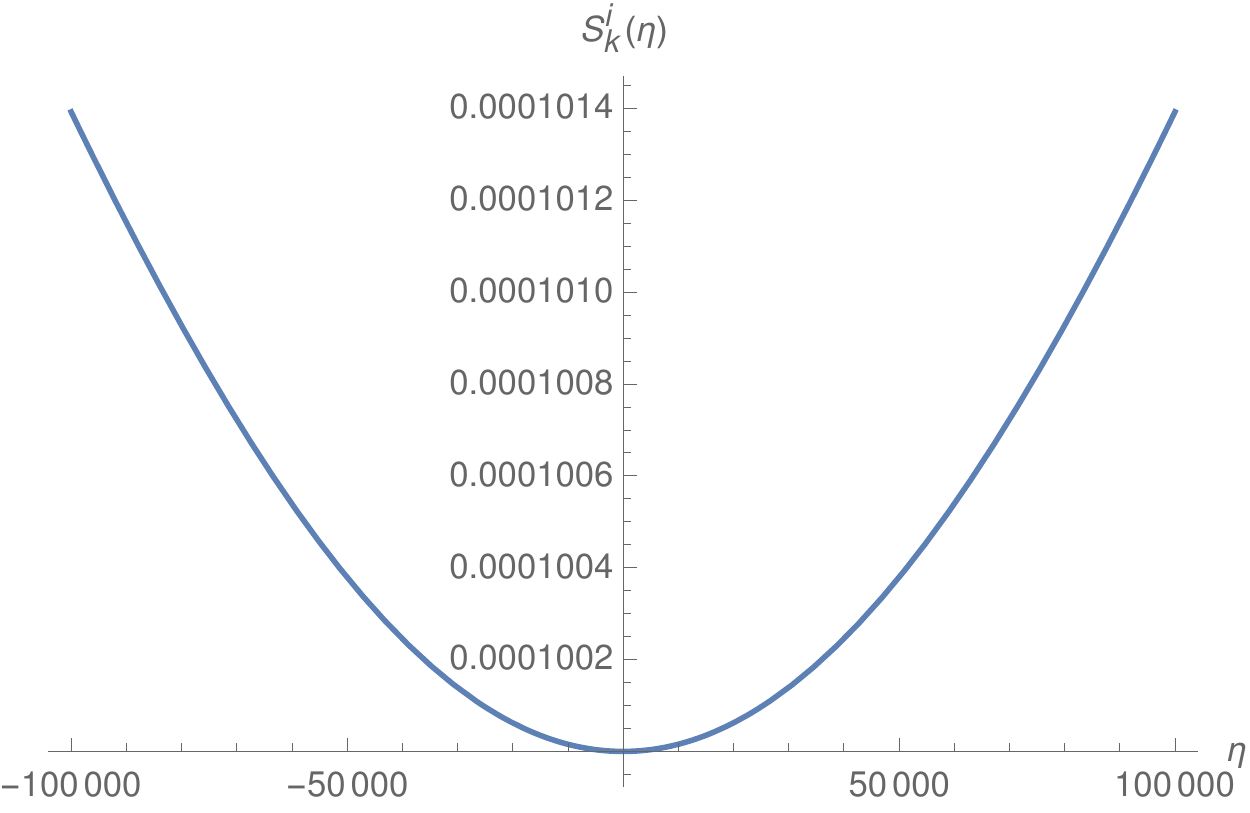}
\caption{Evolution of vector perturbation $S^i_k$ with $\eta$ through symmetric
bounce, where $S^i_k(0)=0.0001$.} 
\label{vs1}
\end{minipage}
\hspace{0.2cm}
\begin{minipage}[b]{0.5\linewidth}
\centering
\includegraphics[scale=.6]{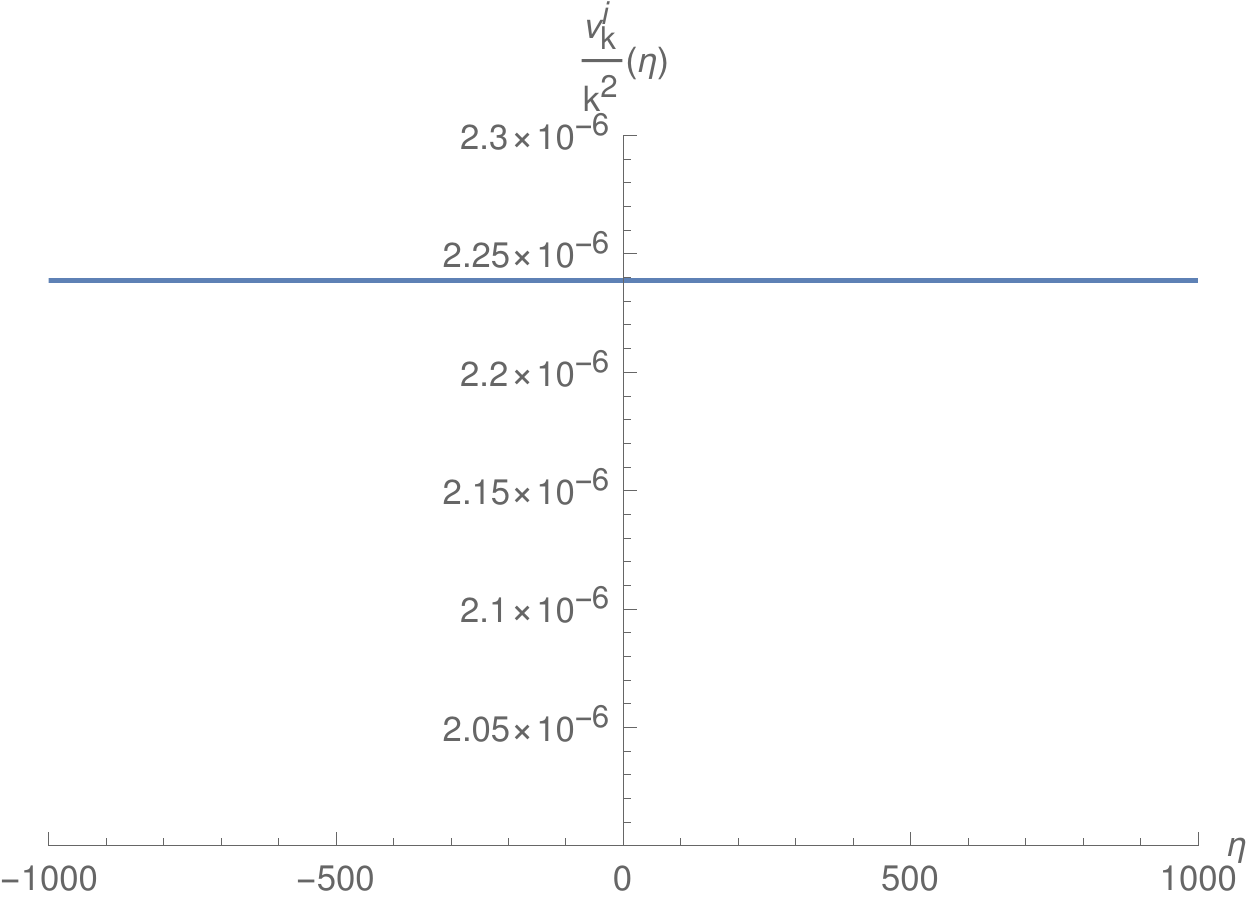}
\caption{Evolution of velocity perturbation $v^i_k$ with $\eta$ through symmetric
bounce. $S^i_k(0)$'s are the same as in the previous plot.}
\label{vs2}
\end{minipage}
\end{figure}

In the Newtonian gauge the solutions of the above equations, in the Fourier space where the subscript $k$ specifies the $k$th mode, are
\begin{equation} \label{vik}
v^i_k = \frac{k^2 F}{2a^2  \kappa (\rho + P)} S^i_k 
\end{equation}
and\\
\begin{equation}\label{sik}
S^i_k = \frac{C^i_k}{a^2 F}
\end{equation}
where $C^{i}_{k} $ is a constant 3-vector. Combining the above equations we get an expression for the Fourier mode of velocity perturbation as:
\begin{equation}\label{vpertf}
v^{i}_{k} \sim  \frac{k^2 C^{i}_{k}}{a^{1-3\omega}}\,.  
\end{equation} 
In GR, in contracting phase of the universe, the vector metric
perturbation increases as the scale-factor decreases. This growth of
perturbations could pose a fundamental threat to the validity of
perturbation theory. But, in $f(R)$ theory, the evolution of vector
perturbation depends on the term $a^2F(R)$. The behavior of $F(R)$ in
general affects the evolution of the vector perturbations in metric
$f(R)$ cosmology. 

In this paper we work with exponential gravity where the form of
$f(R)$ is given in Eq.~(\ref{frform}) where $\alpha= 10^{12}$.  The
metric perturbation $S^i_k$ and velocity perturbation $v^i_k$ for
$k=10^{-10}$, for a radiation dominated universe where $\omega=1/3$,
have been plotted for a symmetric cosmological bounce in the
background in Fig.~\ref{vs1} and Fig.~\ref{vs2}. The conditions used
for the background evolution for the symmetric bounce remains the same
as specified in subsection \ref{jords}. The plots clearly show that in
$f(R)$ cosmology the vector perturbations remains practically the same
through the non-singular bounce. More over the metric vector
perturbation slightly diminishes during the contracting phase where as
in GR the vector perturbation only increases during the contracting
phase. Consequently $f(R)$ cosmology can moderate the growth of vector
perturbations, a fact perhaps noted for the first time in this paper.

In the Einstein frame analysis of vector perturbations, the field equation
is:
\begin{equation}
\tilde{R}^{\mu}_{\nu}-\frac{1}{2}\delta^{\mu}_{\nu} \tilde{R}=\kappa (\tilde{T}^{\mu}_{\nu}+ S^{\mu}_{\nu})\,.
\end{equation}
In the Einstein frame the effective energy-momentum tensor is
$\tilde{T}^{\mu}_{\nu}+ S^{\mu}_{\nu}$ where $S^{\mu}_{\nu}$ is the
energy momentum tensor of a scalar field. $\tilde{T}^{\mu}_{\nu}$ is
the Einstein frame counterpart of ${T}^{\mu}_{\nu}$.  As the scalar
field does not produce any vector perturbations only the fluid
perturbations, coming from $\tilde{T}^{\mu}_{\nu}$ contribute on the
right hand side of the perturbed Einstein equation.  Consequently the vector
perturbation equations in Einstein frame are:
\begin{eqnarray}
  \frac{1}{2\tilde{a}^2} \nabla \tilde{S}^{i}=-\kappa  (\tilde{\rho} +\tilde{P})\tilde{v}^{i}\,,\,\,\,\,{\rm and}\,\,\,\,\,\,
  \partial_{\eta}\left[\tilde{a}^2(\tilde{S}^{i}_{\,\,,j}+ \tilde{S}^{\,\,\,i}_{j,})\right]=0\,,
\end{eqnarray}
where we have used the results from \cite{Battefeld:2004cd}. Solving these equations we get
\begin{equation}
\tilde{S}^{i}_{k} = \frac{\tilde{C}^{i}_{k}}{\tilde{a}^2}\,,
\end{equation}
which shows that $\tilde{S}^{i}_{k}=S^{i}_{k}$ when
$\tilde{C}^i_k=C^i_k$. In the present case we have chosen
$C^i_k=0.00015$. As in the present case the vector perturbations
remain equal in both the conformal frames we do not separately present
the Einstein frame result.  Both the frames produce the same plots and
no singularities are encountered in converting the result from Einstein
frame to the Jordan frame. The present calculation shows that one can
always use the Einstein frame for the calculation for vector
perturbations as the calculations are much easily handled in the
Einstein frame.
\section{Discussion and conclusion}

The present paper deals with metric perturbations in a bouncing
cosmology guided by $f(R)$ theory of gravity. Scalar and vector metric
perturbations during bounce in exponential gravity have been presented
in this paper. The cosmological bounce takes place in the Jordan
frame. The scalar metric perturbations have been calculated in the
longitudinal gauge as this gauge choice remains invariant under a
conformal transformation relating the Jordan frame and the Einstein
frame. In the case of the vector perturbations the metric
perturbations remain the same in both the frames and one can work with
any gauge one wishes to. In the present paper we have worked with the
Newtonian gauge. While the background evaluation of bouncing
cosmologies can efficiently be calculated in the Einstein frame the
perturbations on the metric and fluid variables can also be calculated
in the Einstein frame where we expect the gravitational theory to be
like GR with an extra scalar field. In a previous calculation the
authors tried to calculate the scalar perturbations in a bouncing
scenario using the Einstein frame \cite{Paul:2014cxa} but did not take
into account the fluid perturbations. Hence the earlier calculations
were incomplete. In this paper the full calculation of the scalar
perturbations in the Einstein frame are presented for the first
time. The perturbation calculations are general although in the
present case the results are applied for the particular case of a
cosmological bounce in the Jordan frame.  All the results presented in
this paper is for an exponential $f(R)$ gravity theory which satisfies
the basic stability conditions\footnote{Some general remarks on
  stability of $f(R)$ theories were presented in
  Ref.~\cite{Barrow:1983rx}. In the referred work the authors find out
  the stability of $f(R)$ cosmology by slightly perturbing the
  scale-factor and the issue of metric perturbations was not
  studied.}. The matter content of the universe was always assumed to
be radiation fluid as these is the most general fluid which plays an
important role in the very early universe.

In the present paper the critical role of the Einstein frame in aiding
the calculations of Jordan frame phenomena in $f(R)$ gravity is
studied in full detail. The paper establishes that the Einstein frame
can be a used for the background evolutions for all the cases in the
Jordan frame and also for perturbation evolution of many cases in
$f(R)$ gravity except the particular case of asymmetrical bounce in
the Jordan frame where one cannot transform the scalar metric
perturbation back to the Jordan frame. Although the correspondence of
the background cosmological evolution in the conformal frames, for
$f(R)$ based cosmology, was noted much earlier \cite{Maeda:1988ab,
  Mukhanov:1990me} here one must note that in \cite{Paul:2014cxa} it
was pointed out that for the case of a cosmological bounce in presence
of matter, which satisfies $\rho+P \ge 0$ in the Jordan frame, there
is no corresponding cosmological bounce in the conformally related
Einstein frame when one works with spatially flat FLRW spacetimes.  As
far as cosmological bounce in flat FLRW spacetime is concerned, the
Einstein frame acts like a true auxiliary frame where one can do all
the calculations and then convert the results appropriately in the
Jordan frame. For cosmological perturbations one can also use the
Einstein frame as the auxiliary frame but as far as scalar
perturbations are concerned the conformal correspondence fails for
asymmetric bounces. This particular failure is not a physical problem
as shown in the present paper. The perturbations evolve smoothly in
both the Jordan frame calculation and Einstein frame calculation. The
problem arises in the connecting formulae which connect the Jordan
frame perturbations with the Einstein frame perturbation. The paper
presents the calculations of perturbations separately in both the
conformal frames and then connects the results to show the validity of
the Einstein frame results. The nature of the variation of the fluid
velocity potential in the Jordan frame is also presented in the paper.

The next part of the present paper deals with the evolution of vector
perturbations during a non-singular bounce in $f(R)$ gravity. This
calculation shows that for vector perturbations one can actually use
the Einstein frame for all the cases and the calculations do become
much easier in the Einstein frame. The vector perturbations remains
the same in both the frames and no singularity appears anywhere
in the description of vector perturbations from the Einstein frame. In
GR based cosmological models it was shown that one expects the vector
perturbations to be increasing during the contracting phase. For a
singular bounce the vector perturbations can diverge during the
bounce. In the present paper we have only studied non-singular bounce
in exponential gravity and the results regarding the evolution of
vector perturbations during such a phase are interesting. In this case
the vector perturbations practically remains constant during the
bouncing phase showing stability of the perturbation modes. One can
expect that in the deep expansion phase of the universe this vector
modes do get diluted. There has been studies on magnetic field
generation in the early universe and the nature of vector
perturbations, we expect the specific nature of the evolution of
vector perturbations in the present case can have interesting
consequences for magnetogenesis.

The question of tensor perturbation in general $f(R)$ gravity and in
particular related to bouncing scenarios has not been presented in the
paper. Tensor perturbations will be taken up in a future publication
as there are various interesting issues related to primordial tensor
perturbations which require separate and thorough investigation. The
present work presupposes that the earlier universe (much before the
bouncing time) and the later universe (much later than the bouncing
time) were guided by a theory of gravity like GR which gets
effectively modified to exponential gravity in the high curvature
limit near the bounce. The change over from GR to $f(R)$ is
non-trivial and will require new physics. The perturbation modes
earlier and later than the bounce are expected to follow standard
cosmological dynamics obtained from GR. In this process our main aim
is to show how the perturbation modes evolve during the bouncing
time. It is shown that the perturbation modes can solely be tackled
from the accompanying, conformally connected Einstein frame for most
of the cases. The perturbation modes which have been studied remain
perturbative throughout the bounce, but this does not comprehensively
specify that all the perturbations are stable. The vector
perturbations should be stable for all initial values but it is very
difficult to generalize the statement for scalar perturbations for all
initial values of the perturbations. There may remain some modes with
specific initial conditions which tend to become non-perturbative near
the bounce. In such cases one has to bring in new physics to settle
the issue of perturbative instability.  While our work does not prove
perturbative stability in the most general way it definitely shows how
the stable perturbations evolve near the bounce.


\end{document}